\documentstyle[aps,epsfig,twocolumn]{revtex}

\draft
\newcommand{\pathfigs}{.}

\newcommand{\be}{\begin{equation}}
\newcommand{\ee}{\end{equation}}
\newcommand{\bea}{\begin{eqnarray}}
\newcommand{\eea}{\end{eqnarray}}
\newcommand{\bdm}{\begin{displaymath}}
\newcommand{\edm}{\end{displaymath}}

\newcommand{\bi}{\begin{itemize}}
\newcommand{\ei}{\end{itemize}}

\newcommand{\refkl}[1]{(\ref{#1})}

\newcommand{\abl}[2]{\frac{\partial #1}{\partial #2}}  
 
\newcommand{\ablii}[2]{\frac{\partial^{2} #1}{\partial #2^{2}}} 
 
\newcommand{\erw}[1]{\mbox{$\langle #1 \rangle$}} 

\newcommand{\intd}[1]{\int \!d#1}

\newcommand{\tilrho}{\tilde{\rho}}

\newcommand{\rhomax}{\rho_{\rm max}}

\begin{document}
\pagestyle{myheadings}
\markboth{Treiber/Hennecke/Helbing: Gas-Kinetic-Based Non-Local Traffic Model}
{Treiber/Hennecke/Helbing: Gas-Kinetic-Based Non-Local Traffic Model}




\twocolumn[\hsize\textwidth\columnwidth\hsize\csname @twocolumnfalse\endcsname

\title{Derivation, Properties, and Simulation of a Gas-Kinetic-Based,
Non-Local Traffic Model}
\author{Martin Treiber,
        Ansgar Hennecke,
        and Dirk Helbing}
\address{II. Institute of Theoretical Physics, University of Stuttgart,
         Pfaffenwaldring 57, D-70550 Stuttgart, Germany}

\date{\today}
\maketitle
\vfill
\begin{abstract}
We derive macroscopic traffic equations from specific gas-kinetic equations,
dropping some of the assumptions and approximations
made in previous papers. The resulting
partial differential equations for the vehicle density and
average velocity contain a non-local interaction term which is very favorable
for a fast and robust numerical integration, so that several thousand
freeway kilometers can be simulated in real-time.
The model parameters can be easily calibrated by means of
empirical data. They are directly related to the quantities characterizing
individual driver-vehicle behavior, and their optimal values have
the expected order of magnitude. Therefore, they allow to investigate the 
influences of varying street and weather conditions or freeway
control measures. Simulation results for realistic model parameters
are in good agreement with the diverse 
non-linear dynamical phenomena observed in freeway traffic.  
\end{abstract}
\pacs{05.70.Fh,05.60.+w,47.55.-t,89.40.+k}

] 

\section{Introduction}

Recently, traffic dynamics has become interesting to a rapidly growing
community of physicists. This is not only due to its practical implications
for optimizing freeway traffic, but even more because of the observed
non-equilibrium phase transitions 
\cite{kerner-sync,helb-nat}
and non-linear dynamical phenomena like the formation of
traffic jams 
\cite{kk-93,helb-imp95},
stop-and-go traffic 
\cite{kk-94}, 
and synchronized traffic \cite{kerner-sync,kerner-rehb96-2}.
It seems that all forms of congested traffic have
almost universal properties which are largely
independent of the initial conditions and the spatially averaged density, like
the characteristic outflow from traffic jams of about
$1800\pm 200$ vehicles per kilometer and lane or their typical dissolution 
velocity of about $-15\pm 5$ kilometers per hour
\cite{kerner-rehb96}.
This universality arises from the
highly correlated state of motion produced by traffic
congestions \cite{kerner-chania,helb-CA}.
\par
Whereas classical approaches focussed on reproducing the empirically observed
flow-density relation and the regime of unstable
traffic flow, recent publications pointed out that it is more
important to have traffic models which are able to describe the observed 
spectrum of non-linear phenomena and their characteristic properties
\cite{kk-93,kk-94,kerner-rehb96,bando}.
We think that it would be desireable to develop
models that are consistent with both aspects of empirical data. 
Such models have been proposed lately, including cellular automata models 
\cite{Slow-to-Start,helb-CA} and ``microscopic``
models of driver-vehicle behavior 
\cite{krauss,GFM}, and
the macroscopic model discussed in this paper. 
\par
In order to have meaningful and measurable model parameters, 
we will relate our macroscopic model of freeway traffic to
a ``microscopic'' model of driver vehicle behavior via a gas-kinetic
derivation (cf. Sec.~\ref{themodel}). 
Derivations of this kind have been already proposed in
a number of previous publications 
\cite{prigogine,paveri-fontana,phillips-kin,helb-gas},
but the correct treatment of the most interesting regime of moderate and
high densities remained a problem. In 
\cite{helb-physica,wagner-96,klar-97},
the effect of vehicular space requirements has
been taken into account by a correlation factor reflecting the
increased interaction rate of vehicles (which can now be derived from
simple and plausible arguments, cf. Sec.~\ref{basiceq}).
In deriving the associated macroscopic
equations, different approximations have been suggested, the most harmless
of which was a gradient expansion 
\cite{helb-physica,helb-gas98}.
This lead to 
a viscosity term and some unexpected but essential high-density
corrections containing spatial derivatives of different orders. 
However, the resulting partial differential equations were not very 
suitable for numerical simulations. 
\par
Meanwhile we managed to evaluate the Boltzmann-like gas-kinetic
interaction term exactly (cf. Sec.~\ref{basiceq}). 
Since it turned out that a dynamical variance
equation \cite{helb-imp95,helb-granmat} is
not necessary for a description of the presently known properties
of traffic flows, we replaced it by a constitutive relation
which---corresponding to a quasi-adiabatic approximation---agrees 
with the equilibrium variance. The resulting macroscopic 
traffic equations are coupled non-linear partial differential
equations which can be represented in form of
flux equations with a non-local and anisotropic source term
\cite{helb-flux}.
For this reason, we can
now apply various standard methods for numerical integration. It turns
out that the non-local term has similar smoothing
properties like a viscosity term, but it does not change the 
hyperbolic character of the partial differential equations to a parabolic 
one, and it has more
favorable properties with respect to numerical stability. 
For this reason, our model
allows a robust real-time simulation of freeway stretches up to several
thousand kilometers on a usual PC. 
\par
Compared to previous macroscopic traffic models, the gas-kinetic-based 
traffic model (GKT model)
proposed in the following takes into account the variance of
vehicle velocities, which is basically proportional to the square of
average velocity (cf. Eq.~(\ref{thetaA})), 
but with a density-dependent prefactor that determines
the exact form of the flow-density relation in equilibrium. Moreover,
the ``optimal velocity'' or ``dynamical equilibrium velocity``
$\tilde{V}_{\rm e}$ (cf. Eq. (\ref{Vedyn})), towards which the
average velocity relaxes, depends not only on the local density, but also
on the average velocity, and, even more important, on the density and average
velocity at an ``interaction point'' which is advanced by about the
safe distance. Nevertheless, the equations are structurally related to,
for example, the Kerner-Konh\"auser model \cite{kk-93},
so that we find
many similar non-linear phenomena. This includes the sequence of
stable, linearly unstable,
and metastable regimes 
\cite{kk-94,kerner-dipole}, 
the local breakdown effect
\cite{kerner-ramp}, 
the local cluster effect \cite{kk-94},
and the formation of dipole layers \cite{kerner-dipole} 
at sufficiently large densities.
We also obtain that, in the unstable traffic regime, the resulting 
flow-density relation differs from the equilibrium one 
(lying below the latter, cf. Fig.~\ref{fig_Qtheodots}). 
In addition, we find that the outflow
from traffic jams is independent of the initial conditions and the
spatially averaged density (cf. Fig.~\ref{fig_phase}). 
Moreover, the dissolution velocity of
traffic jams varies only a little with density (cf. Fig.~\ref{fig_vg}). 
Finally, the gas-kinetic-based
model is able to explain the common phenomenon of sychronized congested
traffic \cite{helb-MT-sync}, 
if the inflow at on-ramps is taken into account. 
\par
It turns out that our model can be easily calibrated to the static and
dynamic properties of traffic flow data by a certain systematic
procedure (cf. Sec.~\ref{results_diss}). 
All parameters have a clear interpretation, since they
are related to quantities characterizing the driver vehicle units
like desired velocities or vehicle lengths (cf. Table~\ref{tab_par}). 
At least some of the quantities
like the typical desired velocity or the average 
time headway are directly measurable. Moreover, the optimal parameters
obtained from a calibration to empirical data have the expected
order of magnitude (cf. Table~\ref{tab_par}). 
Therefore, the model allows to investigate the
effect of speed limits, of a larger percentage of trucks, of
bad weather conditions, etc. It will usually be sufficient to 
change the affected parameter values accordingly, instead of
calibrating or even modifying the whole model for every new situation.

\section{The model}
\label{themodel}

\subsection{Underlying Gaskinetic Equation} 
\label{kin_eqs} 
%
Similar to the gas-kinetic derivation of macroscopic
equations for fluids,
we start by formulating a kinetic equation for the 
locally averaged dynamics of driver-vehicles units, which
play the role of the molecules, here.
The kinetic equation describes the evolution
of the coarse-grained phase-space density
\begin{eqnarray}
\label{tilderho}
\tilrho(x,v,t)
  &=& \sum\limits_{\alpha} 
      \intd{t'} \intd{x'} \intd{v'} g(t-t', x-x', v-v') \nonumber \\
  & & \times \ \delta\big(x'-x_{\alpha}(t)\big)
      \ \delta\big(v'-v_{\alpha}(t)\big) \, ,
\end{eqnarray}
denoting the probability density of
finding, at a given time $t$, a vehicle $\alpha$ 
at position $x_{\alpha}$ with
velocity $v_{\alpha}$. In more intuitive terms, 
$\tilde{\rho}$ corresponds to the spatial density
of vehicles per lane times their velocity distribution.
Since the GKT model is an effectively one-lane model
where lane changes and overtaking are only implicitely taken into account,
there is no lane index. 
%
%
The coarse graining is performed by taking local averages over a
weighting function 
$g(t-t', x-x', v-v')$ 
satisfying 
$\intd{t'} \intd{x'} \ \intd{v'} \ g(t-t', x-x', v-v') = 1$,
which is localized in a
microscopically large and macroscopically small neighborhood
around $x$ and in suitable neighborhoods around $t$ and $v$.
The particular choice of $g$ is not relevant for the {\em form} of the
macroscopic equations \cite{note-coarsegraining}. However, the scales
$\Delta t$, $\Delta x$, and $\Delta v$ are meaningful in that they
enter the effective relations for the higher velocity moments like the
variance. 

The information specific to vehicular traffic is contained
in the ``microscopic'' dynamics of individual driver-vehicle units.
In the GKT model, we assume the vehicular dynamics of the 
form \cite{GFM,helb-book}
\begin{eqnarray}
\label{xmic}
\frac{d x_{\alpha}}{dt} &=& v_{\alpha},\\
\label{vmic}
\frac{dv_{\alpha}}{dt} &=& \frac{v_{\alpha}^0 - v_{\alpha}}{\tau_{\alpha}}
                     - \sum\limits_{\beta\neq\alpha} f_{\alpha\beta}.
\end{eqnarray}
The first term on the right-hand side of Eq. \refkl{vmic} 
represents the acceleration of the driver-vehicle unit $\alpha$
to the desired velocity $v_{\alpha}^0$
with an adaptation time of $\tau_{\alpha}$. On empty roads,
this is the only acceleration term.
Notice that $v_{\alpha}^0$ is an intrinsic property of the 
driver-vehicle unit.
The second  term on the right-hand side of Eq. \refkl{vmic} 
represents the braking interaction 
of vehicle $\alpha$ due to  slower vehicles 
$\beta$ in front. It depends mainly on the subjective
minimum safe time headway $T_{\alpha}$ to the car in front
that driver $\alpha$ wants to keep.
The details of the braking interaction will be discussed below.

In general, the parameters $v_{\alpha}^0$, $\tau_{\alpha}$,
and the parameters of
the braking interaction like $T_{\alpha}$
are different for each individual
vehicle $\alpha$.
This could be respected by generalizing $\tilrho$ to a
multi-dimensional phase-space density in a 
phase space spanned by the dimensions $x$, $v$, $\tau$, $v^0$, etc.
This density would express the probability of finding at $x$ a
driver-vehicle unit  with velocity $v$, whose microscopic parameters
are $\tau$, $v^0$, etc.
Paveri-Fontana 
applied this concept
to the extra variable $v^0$ alone, formulating an equation for
$\hat{\rho}(x,v,v^0,t)$ \cite{paveri-fontana},
which was further investigated by Helbing \cite{helb-gas}
and Wagner {\it et al.} \cite{wagner-96}.

In the GKT model, however, we  assume that all
deviations of the individual driving behavior 
from that of the ``average driver'' eventually lead to
fluctuations of the velocity.
For a nearly empty road this is obvious. Then,
the braking term is negligible, and
the distribution of vehicle velocities must converge
to that of their desired velocities.
In dense traffic, there are additional sources of velocity
fluctuations. The time when vehicle $\alpha$
starts to brake in response to a slower vehicle in front depends on 
the individual safe time headway $T_{\alpha}$.
Obviously, driver $\alpha$ brakes later for lower values of $T_{\alpha}$.
Thus, different $T_{\alpha}$ lead to different velocities, 
even if {\it all} other
parameters of the vehicles (in particular the desired velocities)
and all initial conditions are unchanged.
In addition, imperfect driving behavior
such as delayed acceleration or overbraking (i.e., braking more than
necessary in a given situation) contributes to the velocity fluctuations
\cite{helb-physica}.

If we are not interested in
microscopic details, such fluctuations
can be described in a global way by
a fluctuating force in the acceleration equation \refkl{vmic}, 
in analogy to hydrodynamic fluctuations \cite{landau-hyd}.
Thus, we approximate Eq. \refkl{vmic} by
\be
\label{vmicstoch}
\frac{dv_{\alpha}}{dt} = \frac{V_0 - v_{\alpha}}{\tau}
                    - \sum\limits_{\beta\neq\alpha} 
                      \overline{f}_{\alpha\beta}
                    + \xi_{\alpha}(t),
\ee
where 
$V_0 = \erw{v_{\alpha}^0} := \frac{1}{N} \sum_{\alpha} v^0_{ \alpha}$ 
and $1/\tau = \erw{1/\tau_{\alpha}}$
are the averaged microscopic parameters of the accelaration 
term ($N$ denotes the total number of vehicles).
The braking term $\overline{f}_{\alpha\beta}$ is
formulated with averaged parameters like $T = \erw{T_{\alpha}}$ as well.
The fluctuating forces $\xi_{\alpha}(t)$ obey
\be
\label{xi}
\erw{\xi_{\alpha}(t)} = 0, \ \ \
\erw{\xi_{\alpha}(t)\xi_{\beta}(t')} 
       = 2 D \delta_{\alpha\beta} \delta(t-t').  
\ee
The  fluctuation strength $D$
will be determined empirically by comparing the resulting velocity
variance with that obtained from single-vehicle data, see below. 
For low traffic densities, where the interactions can be
neglected, Eq. \refkl{vmicstoch} with \refkl{xi} is an ordinary stochastic
differential equation for $v_{\alpha}$. In the stationary 
limit, it leads to the distribution function 
$w(v_{\alpha}) = (2\pi D\tau)^{-1} \exp[ -(v_{\alpha}-V_0)^2/(2 D \tau)]$.
This means that, in the low-density limit of
negligible interactions, the fluctuating strength $D$ of
the fluctuating forces \refkl{xi} is related to the
velocity variance $\theta = \erw{(v_{\alpha}-V_0)^2}$ by
$\theta = D \tau$ (fluctuation-dissipation relation).
In Sec. \ref{basiceq}, it will be shown
that this relation holds for stationary traffic at all densities.\label{xyz}

%
The equations \refkl{vmicstoch} with \refkl{xi} represent a 
microscopic traffic model on its own. 
It remains, however, to specify the ``microscopic'' braking interactions
$\overline{f}_{\alpha\beta}$.
In real traffic, these interactions depend in
a complicated manner on the own velocity,
and on the distances and velocities of the vehicles in front.
To formulate reasonable assumptions for the GKT model,
we will use averaged quantities like the density
$\rho(x,t) = \intd{v} \tilrho(x,v,t)$, which is justified since
we want to derive a macroscopic model.
As a consequence, however,
a simple purely microscopic equivalent of the GKT model
exists only in some special cases.

Specifically, we make the following 
simplifying assumptions:
%
i) A driver at position $x$ reacts
to the traffic situation
at the advanced ``interaction point''
\be
\label{xa}
x_{\rm a} = x + \gamma (l + T v),
\ee
where $l= 1/\rhomax$ (with maximum density $\rhomax$)
is the average vehicle length
plus the bumper-to-bumper distance kept in standing traffic,
and $\gamma$ is an average anticipation factor with typical values
between 1.0 and 3.0.
In the limit of congested traffic,
the interaction point is $\gamma$ vehicle positions
in front of the actual vehicle position $x$. Notice that, in the limit
of congested traffic and for $\gamma=1$,
assumption i) corresponds to 
classical car-following models \cite{FL,bando,helb-book,GFM}.
ii)
In the light of a mean-field ansatz, the traffic situation
at the interaction point can be described
by  the density and the velocity 
distributions at this place, i.e.,  by the phase-space
density 
$\tilrho(x_{\rm a},v,t)$.
iii) There is a certain percentage 
$p(\rho_{\rm a}) \equiv 1/\chi(\rho_{\rm a})$ of interaction-free space
that allows drivers to approach the respective car in front
before they brake. This percentage is a decreasing function of
the average density $\rho_{\rm a} = \intd{v} \tilrho(x_{\rm a},v,t)$
at the interaction point with $p(0) = 1$, and $p(\rhomax)=0$. 
It is determined by the condition
that, in homogeneous dense traffic, the vehicles follow each other
with a time headway of $T$.
Furthermore, we assume that the probability of undelayed
overtaking by  lane changing
is given by $p$ as well \cite{note-gaps}.
Notice, that the factor $\chi=1/p$ can be interpreted as the increase
of the interaction rate due to the finite space requirements
and positional correlations of vehicles, compared to
point-like objects \cite{helb-physica,wagner-96,klar-97,helb-book}.
iv) If a driver is faster than the velocity $v_{\rm a}$
at the interaction point,
and if he cannot overtake by lane changing
[what happens with probability $(1-p)$],
he reduces his velocity abruptly to $v_{\rm a}$ as soon as the
distance to the interaction point (moving with velocity $v_{\rm a}$)
has decreased by his share
$\Delta x_{\rm free} =  p /\rho_{\rm a}$
of the interaction-free space\cite{note-gamma}.
($1/\rho_{\rm a}$ is the average center-to-center
distance between two vehicles at $x_{\rm a}$).

Now, we formulate the kinetic equation underlying the GKT model.
Taking the time derivative of the definition \refkl{tilderho} of
the phase-space density, and inserting
the microscopic equations
\refkl{xmic} and \refkl{vmicstoch},
gives, by partial integration, 
the kinetic evolution equation for the phase-space 
density \cite{helb-book},
\be
\label{kin}
\abl{\tilrho}{t} + \abl{}{x}(\tilrho v)
             + \abl{}{v} \left[ \tilrho \frac{V_0 - v}{\tau} \right]
    =  \abl{}{v} (\tilrho f_{\rm int})  + \ablii{}{v} (\tilrho D) \, ,
\ee
where the interaction term has the general form
\begin{eqnarray}
\label{f_gen}
f_{\rm int} &=& 
              \tilrho^{-1} 
              \sum\limits_{\alpha} \!\!\sum\limits_{\beta(\neq\alpha)}\!\!
               \intd{t'}\!\! \intd{x'} \!\!\intd{v'}  g(t-t', x-x', v-v') 
                                          \nonumber \\
            & & \times \ \overline{f}_{\alpha\beta} \,
              \delta[x'-x_{\alpha}(t)] \ \delta[v'-v_{\alpha}(t)].
\end{eqnarray}
%

The four assumptions for  the microscopic braking
%
interactions $\overline{f}_{\alpha\beta}$ 
directly result in a  Boltzmann-like interaction with a density-dependent
prefactor $P(\rho)$:
\be
\label{f_sgt}
\abl{}{v} \left( \tilrho f_{\rm int} \right)   
  =  P(\rho) {\cal I}(x,v,t)
\ee
with
\begin{eqnarray}
\label{Bolz}
{\cal I}(x,v,t) =
    \int \limits_{v'>v} \!\! dv' (v'-v) \tilrho(x,v',t) \tilrho(x_{\rm a},v,t) 
                    \nonumber \\
  - \int \limits_{v'<v} \!\! dv' (v-v') \tilrho(x,v,t) \tilrho(x_{\rm a},v',t).
\end{eqnarray}
The first term of \refkl{Bolz}
describes the increase of the phase-space density $\tilrho(x,v,t)$
due to the deceleration of faster vehicles with
velocity $v'>v$ which cannot overtake vehicles at $x_{\rm a}$ driving
with velocity $v$, whereas the second term delineates the decrease of the
phase-space density due to decelerations of vehicles driving
with $v$  which cannot overtake slower vehicles at $x_{\rm a}$ 
driving with $v'<v$.
The prefactor 
\be
\label{Pgen}
P(\rho) = (1-p)\chi = \frac{1}{p} - 1
\ee
is proportional to the probability $(1-p)$ that one cannot
immediately overtake a slower vehicle, and
to the correlation factor $\chi=1/p$ describing the increased interaction
rate due to vehicular space requirements.

In summary, the
kinetic phase-space equation upon which the GKT model is based,
is given by 
\begin{eqnarray}
\label{kin_result}
\abl{\tilrho}{t}  
  & + &   \abl{}{x}(\tilrho v)
        + \abl{}{v} \left[ \tilrho \frac{V_0 - v}{\tau} \right] \nonumber\\ 
  & = &  \left(\frac{1}{p}-1 \right) \left[ 
          \int_{v'>v}\!\! dv' (v'-v) \tilrho(x,v',t) \tilrho(x_{\rm a},v,t) 
                    \right.  \nonumber \\
  &   & - \left. \int_{v'<v}\!\! dv' (v-v') \tilrho(x,v,t) 
  \tilrho(x_{\rm a},v',t) \right] \nonumber\\
  & + & \ablii{}{v} (\tilrho D).
\end{eqnarray}
For $\gamma=1$ and for dense traffic, the
underlying microscopic dynamics
is that of a microscopic, stochastic car-following model.
In this case,
traffic behaves like a 
one-dimensional gas of
inelastic hard ``vehicular molecules'' with anisotropic interactions
whose effective sizes vary with the local density such that
there is a space $\Delta x_{\rm free}(\rho_{\rm a})$ between the molecules.

\subsection{ Derivation of the Basic Equations}
\label{basiceq} 
%
Following the standard procedure summarized in 
Refs.~\cite{helb-book,helb-physica}, 
we derive from the kinetic equation \refkl{kin_result}
macroscopic equations for the lowest velocity moments.
In particular, we are interested in
the dynamics of the
macroscopic vehicle density $\rho(x,t)$ per lane and the 
average velocity $V(x,t)$
defined by
\begin{eqnarray}
\rho(x,t)                &=&           \int\limits^{\infty}_0 dv \
                                       \tilrho(x,v,t),      \\
V(x,t) \equiv \erw{v}    &=& \rho^{-1} \int\limits^{\infty}_0 dv \ v   
                                       \tilrho(x,v,t).
\end{eqnarray}
As usual, one obtains an infinite hierarchy of equations where
that for the $n$th moment depends on the $(n+1)$st moment.
In particular, the macroscopic density equation depends on $V$,
and the macroscopic equation for $V$
on the variance
\be
\label{thetagen}
\theta(x,t) \equiv \erw{(v-V)^2} = \rho^{-1} \intd{v} \ (v-V)^2\tilrho(x,v,t).
\ee
In the GKT model, we close the hierarchy by two assumptions. 
First, we assume that the variance $\theta$
is a function of density and average velocity.
Second, we asume that the
phase-space density is locally associated with a 
Gaussian velocity distribution
\be
\label{gauss}
\tilrho(x,v,t) 
                = \rho(x,t) \frac{ e^{-[v-V(x,t)]^2 / [2\theta(x,t)]} }
                                 {\sqrt{2\pi\theta(x,t)}}.
\ee
This is well compatible with 
empirical velocity distributions obtained from single-vehicle data
\cite{helb-book,helb-emp97-2}, at least, 
if the percentage of trucks is negligible \cite{note-bimodal}.
Ansatz \refkl{gauss} is also consistent with
the fluctuating force \refkl{xi}
in the microscopic Eq. \refkl{vmicstoch}.
A more general ansatz taking into account small deviations from
local equilibrium can be found in \cite{helb-granmat}.

Multiplying the phase-space equation \refkl{kin_result} with 1 or $v$,
respectively,
and integrating over $v$
leads, after straightforward but lengthy
calculations, to
\begin{equation}
\label{eqrhoderiv} 
 \frac{\partial \rho}{\partial t} + 
 \frac{\partial (\rho V)}{\partial x} = 0,
\end{equation}
\begin{eqnarray}
\label{eqVderiv}
  \left( \frac{\partial}{\partial t} 
 + V \frac{\partial}{\partial x} \right) V
   &=& 
 - \frac{1}{\rho} \frac{\partial (\rho\theta)}{\partial x}
       + \frac{V_0-V}{\tau} \nonumber \\
       && -  \frac{P(\rho_{\rm a}) \rho_{\rm a} (\theta+\theta_{\rm a})}{2} B(\delta_V),
\end{eqnarray}
where we used the notation $f_{\rm a}(x,t) \equiv f(x_{\rm a},t)$ 
with $f\in\{\rho,V,\theta\}$.
It turned out, that the approximation of the sum
$(\theta+\theta_{\rm a})/2 \approx \theta$
leads only to negligible quantitative changes
\cite{note-approxtheta}. On the other hand,
the approximation simplifies
the velocity equation considerably, so we will adopt it henceforth.
The monotonically increasing macroscopic interaction term
\begin{equation}  
\label{B}
B(\delta_V) =   2 \left[ 
    \delta_V \frac{\mbox{e}^{-\delta_V^2/2}}{\sqrt{2\pi}}
           + (1+\delta_V^2) 
    \int_{-\infty}^{\delta_V} dy \, \frac{\mbox{e}^{-y^2/2}}{\sqrt{2\pi}}
                 \right]
\end{equation}
describes the dependence of the braking interaction on
the dimensionless velocity difference 
$\delta_V = (V-V_{\rm a})/\sqrt{\theta+\theta_{\rm a}}$.
For $\gamma=1$, the macroscopic interaction term can be 
easily understood
by the underlying microscopic dynamics of the GKT model.
If a vehicle at location $x$ with  velocity 
$v$ is faster than one at $x_{\rm a}$ with velocity $v_{\rm a}$ 
(i.e. $\delta v = v-v_{\rm a} > 0$), it approaches 
the  car in front within the time 
$\Delta t = \Delta x_{\rm free}/\delta v$,
where $\Delta x_{\rm free} = p/\rho_{\rm a}$
is the average interaction-free distance 
$\Delta x_{\rm free} = p/\rho_{\rm a}$ of a car.
Then, if it cannot overtake immediately, which would happen with
probability $(1-p)$, it abruptly reduces the velocity by
$\delta v$. The resulting ensemble-averaged deceleration is
\begin{equation}
\label{erwdvdt}
\langle \delta v/\Delta t \rangle
  = - \frac{(1-p)}{p}
  \rho_{\rm a} \int^{\infty}_0 d (\delta v) \ (\delta v)^2 w(\delta v) \, .
\end{equation}
If $v$ and $v_{\rm a}$ are uncorrelated and 
Gaussian distributed [cf. Eq. \refkl{gauss}],
with expectation values $V$, $V_{\rm a}$ and variances $\theta$, 
$\theta_{\rm a}$, respectively, the distribution
function $w(\delta v)$ of the velocity difference
is also a Gaussian, with expectation value $V-V_{\rm a}
= \sqrt{\theta+\theta_{\rm a}}\, \delta_V$ and variance $(\theta+\theta_{\rm a})$.
Evaluating integral \refkl{erwdvdt}
yields $\langle \delta v/\Delta t \rangle = 
 - \frac{1}{2} P \rho_{\rm a} (\theta+\theta_{\rm a}) B(\delta_V)$, i.e.,
the macroscopic braking term in Eq. \refkl{eqVderiv}.

The assumption of a Gaussian velocity distribution alone
would close the system after the variance equation, which
can also be derived from the kinetic equation \refkl{kin_result}
\cite{helb-granmat,helb-flux}.
Since a dynamic variance equation is not necessary for
the description of known traffic instabilities,
we close the system already after the velocity equation
and assume for the variance the 
local equilibrium value $\theta = D \tau$ of
the variance equation \cite{helb-granmat,helb-flux}.
Notice that this relation for the
variance is the same relation as derived in Sec.~\ref{xyz}
for low densities.

To complete the derivation of the GKT equations,
we have to specify the ``constitutive relation''
for the quasi-adiabatically eliminated variance as a function of $\rho$
and $V$, and the relation for the
dimensionless correlation prefactor $P(\rho)$.
The empirical data suggest that the variance
(and thus $D$) is a density-dependent fraction $A(\rho)$ of
the squared velocity,
\be
\label{thetaA}
\theta = A(\rho) V^2,
\ee
and that the variance prefactor $A$ is higher in congested traffic
than in free traffic.
For qualitative considerations, $A$ can be chosen to be constant.
In the following, however, 
we approximate the empirical data by the Fermi function
\be
\label{A}
A(\rho) = A_0 + \Delta A \left[ 
       \tanh \left( \frac{\rho-\rho_{\rm c}}{\Delta \rho} \right) + 1
                 \right],
\ee
where $A_0$ and $A_0 + 2 \Delta A$ are about the variance prefactors
for free and congested traffic, respectively, 
$\rho_{\rm c}$ is of the order of the critical density
for the transition from free
to congested traffic, and $\Delta\rho$ denotes the width of the transition.

Now, we determine the correlation function $P(\rho)$
by imposing that the time headways in dense, homogeneous traffic
are given by $T$.
Solving Eq. \refkl{eqVderiv} for stationary and homogeneous traffic
of density $\rho$ leads to the equilibrium velocity-density relation\label{eq}
\be
\label{Ve}
V_{\rm e}(\rho) = \frac{\tilde{V}^2}{2V_0}
       \left( - 1 \pm \sqrt{1 + \frac{4 V_0^2}{\tilde{V}^2}}
       \right)
\ee
with
\be
\label{tilV}
\tilde{V} = \sqrt{\frac{V_0}{\tau \rho A(\rho) P(\rho)}}.
\ee
This also determines the equilibrium traffic flow per lane by
\be
\label{Qe}
Q_{\rm e}(\rho) = \rho V_{\rm e}(\rho).
\ee
In the limit of high-densities, $(1-\rho/\rhomax) \ll 1$ (or $V_{\rm e} \ll
V_0$), 
this reduces to $V_{\rm e} = \tilde{V}$.
On the other hand, time headways of $T$ and average gaps
of $s = (1/\rho - 1/\rhomax)$ between the vehicles
correspond to
a velocity $V_{\rm T}(\rho) = s/T = (1/\rho - 1/\rhomax)/T$.
Demanding  $V_{\rm e}=V_{\rm T}$ for high densities
leads to
\be
\label{P}
P(\rho) = \frac{V_0 \rho T^2}
         {\tau A(\rhomax) (1-\rho/\rho_{\rm max})^2}.
\ee
This expression is consistent also in the other limit
of homogeneous traffic with very low density. 
With Eq. \refkl{P}, the macroscopic braking term
$-P\rho\theta$ of Eq. \refkl{eqVderiv} for homogeneous traffic is
proportional to $\rho^2$, in accordance with intuition:
The rate of encountering a slower vehicle is  proportional to $\rho$.
Furthermore, 
the probability that one cannot overtake
immediately by lane-changing when a slower vehicle is
encountered, is proportional to $\rho$ as well, resulting in
a proportionality to $\rho^2$ at low densities.
The interpretation of $p=(P+1)^{-1}$ as the percentage
$\rho \, \Delta x_{\rm free}$ of free space is also consistent  
for both limiting cases.
For $\rho\to\rhomax$, one has $\Delta x_{\rm free} \to 0$.
For $\rho/\rhomax \ll 1$, one obtains 
$\Delta x_{\rm free} - \Delta x = V_0 T^2/ (\tau A_0)$.
This means, vehicles on a nearly empty road would 
react to other vehicles in front (mostly by lane-changing), if these
vehicles were closer than $V_0 T^2/(\tau A_0)$. This is the
net safety distance $V_0 T$ times a factor $T/(A_0\tau)$ that is of
order unity (see Table \ref{tab_par}).

\subsection{Discussion of the Model}
\label{diss_model}

For convenience, let us summarize
the equations of the GKT model.
The traffic density and average velocity evolve,
in absence of on- and off-ramps
according to
\begin{equation}
\label{eqrho} 
 \frac{\partial \rho}{\partial t} + 
 \frac{\partial (\rho V)}{\partial x} = 0,
\end{equation}
\begin{eqnarray}
\label{eqV}
  \left( \frac{\partial}{\partial t} 
 + V \frac{\partial}{\partial x} \right) && V
   = 
       - \frac{1}{\rho} \frac{\partial (\rho A V^2)}{\partial x}
         + \frac{V_0-V}{\tau} \nonumber \\
    && -  \frac{V_0 A(\rho)} {\tau A(\rhomax)}
          \left( \frac{\rho_{\rm a} T V}{1-\rho_{\rm a}/\rhomax} \right)^2
          B(\delta_V),
\end{eqnarray}
where $B(\delta_V)$ is given by Eq. \refkl{B}.
(For a generalization to cases with on- and off- ramps 
see Ref. \cite{helb-MT-sync}.)
$A(\rho)$ is the measured or assumed
variance in units of the squared velocity,
for which we use  relation \refkl{A} throughout this paper.

The density equation (\ref{eqrho}) is just
a one-dimensional continuity equation reflecting the 
conservation of the number of vehicles. 
Thus, the temporal
change $\partial \rho/\partial t$ of the vehicle density is just given
by the negative gradient $-\partial Q/\partial x$ of the
lane-averaged traffic flow $Q = \rho V$.  

The first term on the rhs of Eq.~(\ref{eqV}) 
is the gradient of the ``traffic pressure'' $\rho\theta = \rho A V^2$.
It describes the kinematic dispersion of
the macroscopic velocity in inhomogeneous traffic as a consequence
of the finite velocity variance. For example, the macroscopic velocity
in front of a small  vehicle cluster will increase {\it even if no
individual vehicle accelerates}, because the faster
cars will leave the cluster behind. 
The kinematic dispersion also leads to a smooth density
profile at the dissolution front between congested
traffic and an empty road, as it occurs 
when a road blockage at $x_0$
is removed at a time $t_0$. In this case, the first vehicles can all
accelerate to their respective desired velocities. Thus, after 
sufficiently long times, 
the high-speed tail of the distribution 
of desired velocities
translates into a distribution 
of vehicle positions.

The second term of Eq. \refkl{eqV} denotes the acceleration towards
the (traffic-independent) average desired velocity $V_0$ 
of the drivers with a relaxation time $\tau$.
Individual variations of the desired velocity are accounted for by
the finite velocity variance.

The third term of Eq.~(\ref{eqV}) models braking
in response to the traffic situation
at the advanced ``interaction point'' 
$x_{\rm a} = x + \gamma( 1/\rho_{\rm max} + T V)$.
The braking deceleration increases coulomb-like with decreasing 
gap $(1/\rho_{\rm a} - 1/\rho_{\rm max})$ to the car in front
($1/\rho_{\rm a}$ being the average distance between successive vehicle
positions and $1/\rho_{\rm max}$ the minimum vehicle distance).
In homogeneous dense traffic, the acceleration and braking terms 
compensate for each other at about the safe distance. 
In general, the deceleration tendency depends also on the
velocity difference to the traffic at the interaction point,
characterized by the ``Boltzmann factor'' $B(\delta_V)$.
In homogeneous traffic, we have $B(0)=1$.
In the limiting case
$\delta_V \gg 0$ (where the preceding cars are much slower), it follows
$B(\delta_V) = 2 \delta_V^2$. If, in contrast, the preceding cars are
much faster (i.e. $\delta_V \ll 0$), we have
$B(\delta_V) \approx 0$. That is, since the distance is increasing, then, 
the vehicle will not brake, 
even if its headway is smaller than the safe distance.

The main difference with respect to  other macroscopic traffic models 
is the non-local character of  the braking term, which we obtained by
derivation from realistic assumptions of
driving behavior.
The non-locality has very favorable  properties
with respect to the robustness of numerical
integration methods and their integration speed.
It has smoothing properties like
the viscosity term used in the Kerner-Konh{\"a}user model
\cite{kk-93,kk-94}, but its effect is anisotropic. There is
no smoothing in forward direction, which would
imply that cars would react on density or velocity gradients of the
vehicles behind them.

The GKT model fits into the general class of macroscopic traffic
models \cite{helb-physica,helb-book} defined by the continuity equation 
\refkl{eqrho} and the velocity equation
\be
\label{eqVgen}
  \left( \frac{\partial}{\partial t} 
 + V \frac{\partial}{\partial x} \right) V 
     =  - \frac{1}{\rho} \frac{\partial {\cal P}}{\partial x}
         + \frac{1}{\tau} 
  [\tilde{V}_{\rm e}(\rho,V,\rho_{\rm a},V_{\rm a}) - V].
\ee
In the GKT model, the ``traffic pressure'' ${\cal P}$ is given by
\be
\label{calP}
{\cal P} = \rho\theta = \rho A(\rho) V^2,
\ee
and the ``dynamical equilibrium velocity'',
towards which the
average velocity relaxes in the actual traffic situation, 
is
%
\begin{eqnarray}
\label{Vedyn}
\tilde{V}_{\rm e}(\rho,&& V,\rho_{\rm a},V_{\rm a}) \nonumber \\
    &&= V_0\left[
     1 - \frac{A(\rho)}{2 A(\rhomax)}
          \left( \frac{\rho_{\rm a} T V}{1-\rho_{\rm a}/\rhomax} \right)^2
          B(\delta_V) \right].
\end{eqnarray}
In contrast to other macroscopic models 
\cite{payne,phillips-kin,kk-93} belonging to 
the class defined by Eq. \refkl{eqVgen},
the ``dynamical equilibrium velocity'' depends on $\rho$ {\em and} $V$ at
{\em two} different locations, thus introducing the non-locality.

\begin{figure}
\begin{center}
  \includegraphics[width=100\unitlength]{\pathfigs/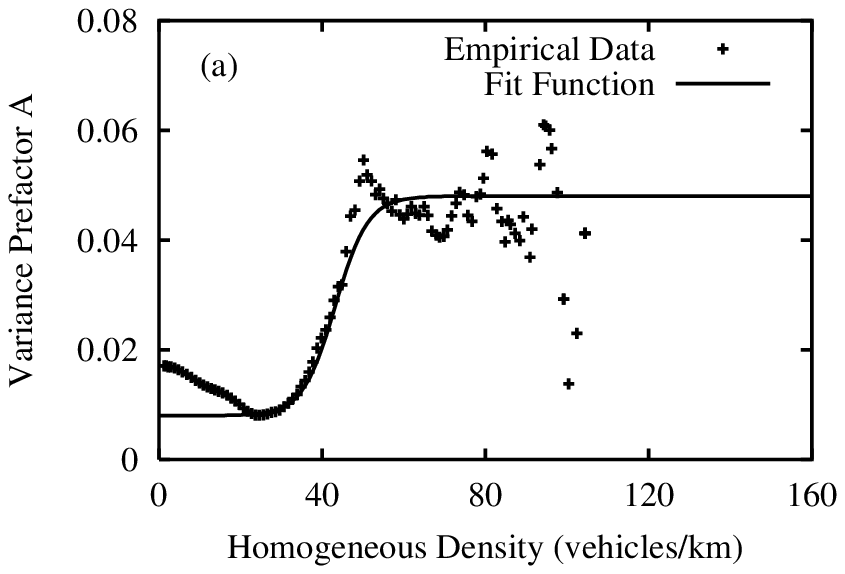}
  \\[0mm]
  \includegraphics[width=100\unitlength]{\pathfigs/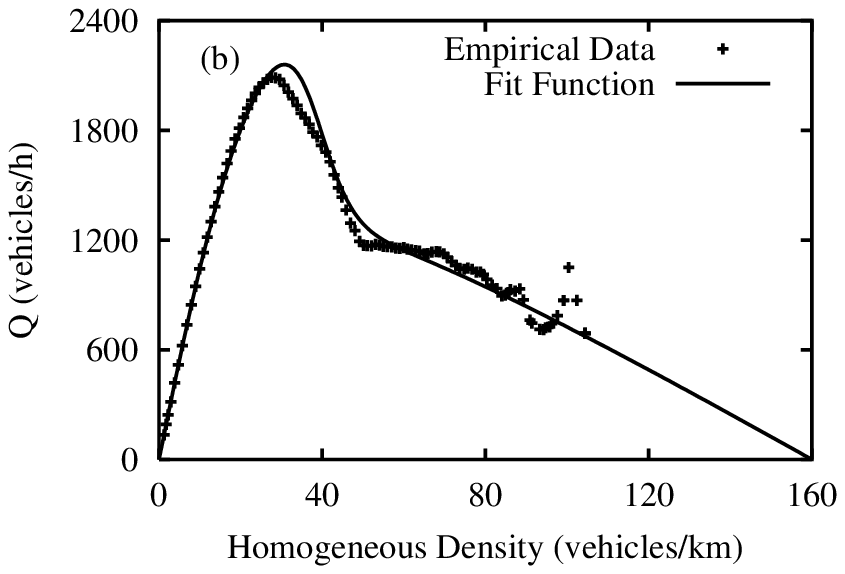}
\end{center}

\caption[]{ \label{fig_AQemp}
Comparison of (a) the density-dependent
relative variance  in units of the squared average velocity,
and (b) the equilibrium flow-density relation
\protect\refkl{Qe} in the GKT model 
(solid lines) with empirical data (crosses).
The empirical data were obtained from single-vehicle data of the Dutch
motorway A9 by averaging over one-minute intervals (see main text).
}
\end{figure}

The five parameters of the GKT model, listed in Table
\ref{tab_par}, are all intuitive. 
Three of them,
$V_0$, $T$, and $\rhomax$,
can be directly determined by fitting the equilibrium
flow density relation \refkl{Qe} of the GKT model to
measured flow-density data,
cf. Fig. \ref{fig_AQemp}.
The desired velocity
$V_0$ is determined by fitting the data
at low densities by a straight line $Q(\rho) = V_0\rho$.
The safe time headway $T$ and the maximum density
$\rhomax$ are determined by fitting the data at high densities by
a straight line crossing the abszissa at $\rhomax$ and
identifying this line with  $Q(\rho) = (1-\rho/\rhomax)/T$. 
The ensuing average distance  $1/\rhomax$ of two cars in 
standing traffic must be consistent with the average length of the
vehicles plus a minimal bumper-to bumper distance kept,
which is  about 1.5\,m. As 
real traffic is stable at very low and very high densities,
the above procedure of comparing
the measured data in these density ranges
with the {\it equilibrium} curve of the GKT model is justified.
At intermediate densities, the equilibrium curve of the model
lies somewhat above the data (Fig. \ref{fig_AQemp}). However,
in Sec. \ref{results} it will be shown 
that homogeneous traffic is unstable in this density range, and that
the averaged  {\it dynamic} traffic flow in the GKT model
is below the equilibrium curve as well.

The remaining parameters $\tau$ and $\gamma$
can be systematically calibrated by means of
the dynamic properties. This will be
discussed in Sec. \ref{conclusion}. 
Table \ref{tab_par} lists the numerical values 
resulting from a fit to traffic data of the Dutch motorway A9.
If not stated otherwise, we used these values
in the numerical simulations of Sec. \ref{results}, referring to them as
``standard parameter set''.

Notice that all parameters have realistic values. In particular, this
holds for $\tau$ which,
for $V_0$ = 158 km/h, would have the meaning of the acceleration
time from 0 to 100 km/h. Furthermore, $V_0/\tau$
is limited to the average maximum acceleration of vehicles on
a free road starting with zero velocity. For these reasons, a relaxation
time $\tau \approx 35$\,s is sensible for freeway traffic. 
(For city traffic, $\tau$ is shorter).

The value $T=1.8$\,s for the safe time headway
is consistent with the rule ``distance (in m) should not be 
less than half the velocity
(in km/h)'' suggested by German authorities.
For other data, however, we  often find
that a somewhat smaller time headway gives a better fit.

Since the model parameters are meaningful, it is 
simple
to model changes of the traffic dynamics caused by external effects
like environmental influences.
For example, a speed limit would be conidered by decreasing $V_0$.
Bad weather conditions leading to more defensive driving 
would be characterized by an increased time headway $T$ and a lower
value of $V_0$ (plus a reduction of $\gamma$, if there is heavy fog).
In rush-hour traffic, it is plausible to assume a higher percentage
of experienced drivers than in holiday traffic, which would correspond to
a higher $\gamma$.
Effects like a varying distribution of vehicle types
can be modelled as well.
For example, a higher proportion of trucks would lead to a decrease of
$V_0$ and $\rhomax$,
but also to an increased value of  $\tau$.


Finally, we compare the macroscopic GKT model with direct simulations of
microscopic models of the form \refkl{vmicstoch}. 
While the microscopic model is stochastic, the deterministic
GKT model includes the stochasticity of real traffic by the constitutive
relation \refkl{thetaA} for the velocity variance.
Therefore, the GKT describes {\em macroscopic} effects of
fluctuations like kinematic dispersion.
The additional information of individual fluctuations
contained in microscopic models seems not to be of practical relevance,
since empirical traffic data are typically available
as one-minute averages, i.e. in terms of  macroscopic quantities.
\par
In contrast to microscopic models,
the GKT model is an effectively one-lane model and
treats overtaking and the associated lane-changing
manoeuvres in an overall way.
A microscopic model would
need additional assumptions and new parameters  for the lane-changing
decisions
as well as additional assumptions about the population of vehicles, e.g.,
the distribution of desired velocities.
Moreover, the macroscopic model can be generalized 
to simulate on-ramps, off-ramps, and lane closings,
simply by adding source and sink terms to the
macroscopic density and velocity equations \refkl{eqrho} and \refkl{eqV}
\cite{kerner-ramp,helb-MT-sync}.
In microscopic models this would require the treatment of
lane changes from dead-end
lanes, which is a particularly difficult problem.
\par
Finally, apart from very low densities, 
the numeric performance of simulations with the
GKT model is far superior to corresponding microscopic simulations.
This is achieved partly by using lookup tables for the functions
$A(\rho)$ and $B(\delta_V)$ and by applying explicit integration 
schemes \cite{Numer}. In addition,
the GKT model has only one density and velocity field variable,
and its computational speed (measured in terms of the
length of the road sections that can be
simulated real-time) is independent of the density and the
number of lanes.


\subsection{Dimensionless Form of the GKT Equations}
\label{dimless}

By reformulating the GKT model in dimensionless variables,
the number of model parameters can be reduced by two.
We measure times in units of $\tau$ and distances in units of
$1/\rhomax$ 
by introducing
$t' = t/\tau$ and $x' = \rhomax x$. The dependent variables
$\rho' = \rho/\rhomax$, $V' = V\tau\rhomax$,
$\theta' = \theta\tau^2 \rhomax^2$, etc. are scaled accordingly.
This leads to the scaled GKT equations
\begin{equation}
\label{eqrhoscal} 
 \frac{\partial \rho'}{\partial t'} + 
 \frac{\partial (\rho' V')}{\partial x'} 
 = 0
\end{equation}
and
\begin{eqnarray}
\label{eqVscal}
  \left( \frac{\partial}{\partial t'} 
       + V' \frac{\partial}{\partial x'} \right) V'
   &=& 
       - \frac{1}{\rho'} \abl{}{x'} (\rho'\theta')
       + (V_0'-V') \nonumber \\
    && - P'A'(\rho') \frac{(\rho'_{\rm a} V')^2}
          {(1-\rho'_{\rm a})^2} B(\delta_V) \, ,
\end{eqnarray}
where 
\be
\label{Ascal}
A'(\rho') =  \frac{A(\rhomax \rho')}{A(\rhomax)} 
\ee
is of order unity,
and the Boltzmann term $B(\delta_V)$ (i.e., Eq. \refkl{B}
in scaled variables), remains
unchanged.
The remaining dimensionless parameters are
the scaled desired velocity
\begin{equation}
\label{V0dash}
V_0' = \rho_{\rm max} \tau V_0
\end{equation}
and the scaled cross-section
\begin{equation}
P' = \frac{\rho_{\rm max} V_0 T^2}{\tau}
   = V_0' \left( \frac{T}{\tau}\right)^2,
\end{equation}
in addition to the anticipation factor
$\gamma$ from the unscaled equations.

The parameter $V_0'$, 
with a numerical value of 171.4
for the standard parameter set,
has some analogies to the
Reynolds number in the Navier--Stokes equations for normal fluids.
Assuming that typical velocities are proportional to the
desired velocities, typical densities proportional to $\rhomax$, and
typical length scales proportional to $1/\rhomax$, 
this can be seen by observing that the 
magnitude of the destabilizing advection,
pressure, and braking terms in the unscaled 
velocity equation  \refkl{eqV} is proportional to
$\rhomax V_0^2$, while the
stabilizing relaxation term is 
proportional to $ V_0/\tau$.  So, the ratio
between the destabilizing ``kinetic'' terms and the stabilizing
relaxation term is proportional to $\rhomax V_0\tau = V_0'$.
As will be shown in Sec.~\ref{stab},
homogeneous traffic can become unstable, if a certain ``critical''
value  $V'_{\rm c}$ is exceeded.

The scaled cross-section $P'$, with a 
numerical value of 0.453 for the standard parameter set,
gives the ratio between the interaction term
and the ``kinetic'' advection and pressure terms.
In analogy to the Prandtl number of thermal convection 
in a simple fluid heated from
below \cite{cross-hohenberg},
it depends on the ratio of the two relevant
time scales $T$ and $\tau$ of the system.

As in the unscaled equations, 
the parameter $\gamma$ characterizes
the sensitivity in the braking interactions to spatial
changes of the velocity, compared to the sensitivity
for changes of the density.

\section{Results} 
\label{results}

\subsection{Homogeneous Traffic} 
\label{hom_traff}

In homogeneous and stationary traffic,
the GKT equations \refkl{eqrho} and \refkl{eqV}
reduce to $\rho = $ const. and relation
\refkl{Ve} for the equilibrium velocity
$V_{\rm e}(\rho)$. Notice that $V_{\rm e}(\rho)$
and thus the equilibrium
flow $Q_{\rm e} = \rho V_{\rm e}(\rho)$ is a function of the
model parameters $V_0$, $T$, and $\rhomax$, but
does not depend on $\tau$ and $\gamma$.

We determined
the constants in the relation \refkl{A} for 
the variance prefactor $A(\rho)$ 
and the model parameters $V_0$, $T$, and $\rhomax$,
by fitting them to empirical data of the Dutch two-lane
motorway A9 from Haarlem to Amsterdam.
The empirical data are based on one-minute values for the number $n_t$
of passing vehicles, their average velocity $V_t$, and velocity variance
$\theta_t$, which were determined from single-vehicle data. The
corresponding flow is then given by $Q_t=n_t/2$ per minute and lane,
and the density per lane by $Q_t/V_t$.
Such sets of one-minute values were sampled over two periods from Monday 
to Friday (October 10, 1994 through October 14,
1994 and October 31, 1994 through November 4, 1994) for the right 
and left lane at nine subsequent measuring cross-sections 
(distributed over a stretch of 8.6 km length; see Ref.~\cite{Ill} for
an illustration).
To obtain empirical quantities as a function of density, we averaged
over all sets with a density between
$\rho-\Delta\rho$ and $\rho+\Delta\rho$ with $\Delta\rho=$1 vehicle/km.

Figure \ref{fig_AQemp}(a) shows the variance prefactor 
Eq. \refkl{A} for the fitted values
\begin{eqnarray} 
 \rho_{\rm c} &=& 0.27 \rhomax , \nonumber \\
 \Delta \rho &=& 0.05 \rhomax , \nonumber \\
 A_0 &=& 0.008 , \mbox{ and} \nonumber \\ 
\Delta A &=& 2.5 A_0 .
\end{eqnarray}
Throughout the paper, we will use these values.

Figure \ref{fig_AQemp}(b) depicts the flow-density relation
for the fitted values of $V_0$, $T$, and $\rhomax$
given in Table \ref{tab_par}. These values resulted
from the systematic procedure
described in Sec.~\ref{results_diss}.
Both the empirical variance-density relation and the 
flow-density relation are well reproduced by introducing
{\it one} fit function $A(\rho)$ only. 
With a constant value for $A$, the sharp increase of the variance
prefactor at a density of about 40 vehicles/km
and the sharp decrease of the velocity related to it could not be
obtained by variation of $V_0$, $T$, and $\rhomax$. 
Rather, this correlation is an intrinsic property of the model and
follows from the proportionality of the braking interaction to the
variance. Notice, that the deviation of the assumed variance prefactor 
from the data at very low densities could be easily removed by a more complicated
function $A(\rho)$. However, this correction would not impair the flow-density
relation, because the interaction is negligible for these densities.
For the same reason, the correction would not change the dynamics,
justifying the choice of the simple functional dependence \refkl{A}.

\begin{figure}
\begin{center}
   \includegraphics[width=90\unitlength]{\pathfigs/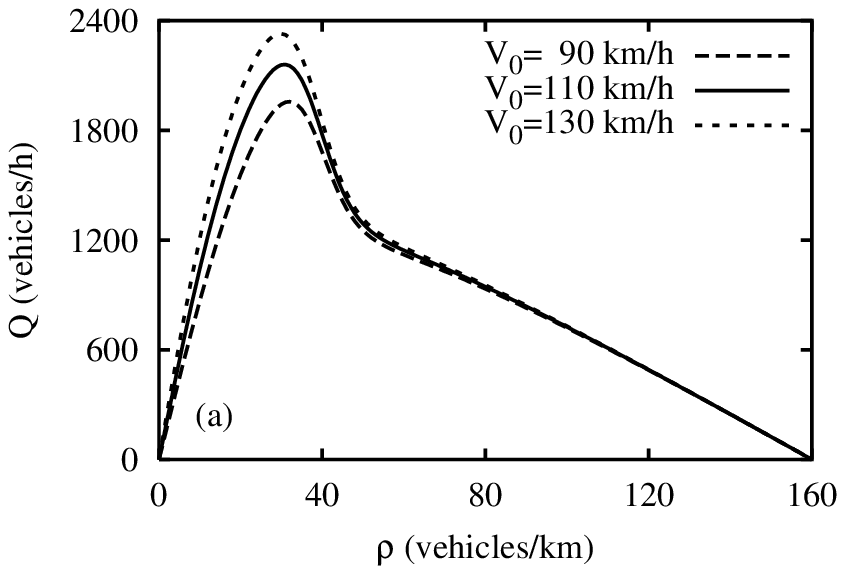} \\[-0mm]
   \includegraphics[width=90\unitlength]{\pathfigs/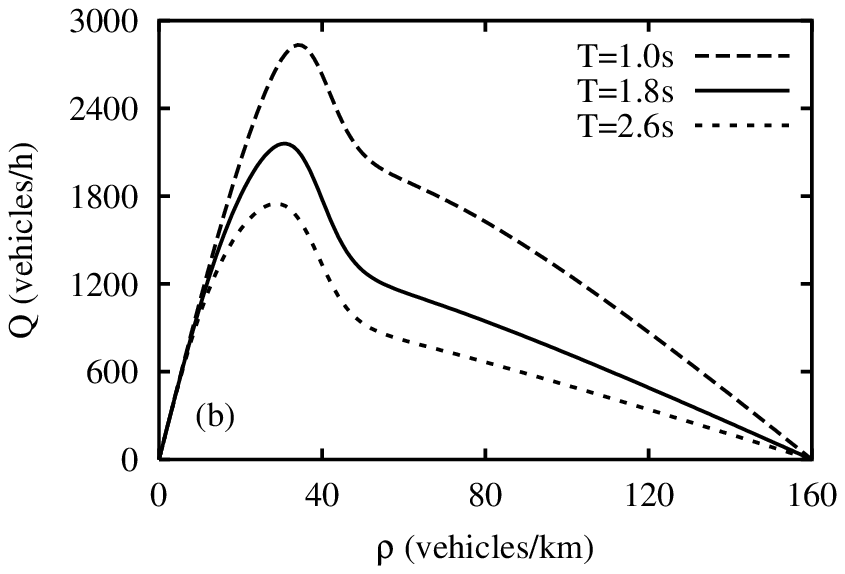} \\[-0mm]

   \includegraphics[width=90\unitlength]{\pathfigs/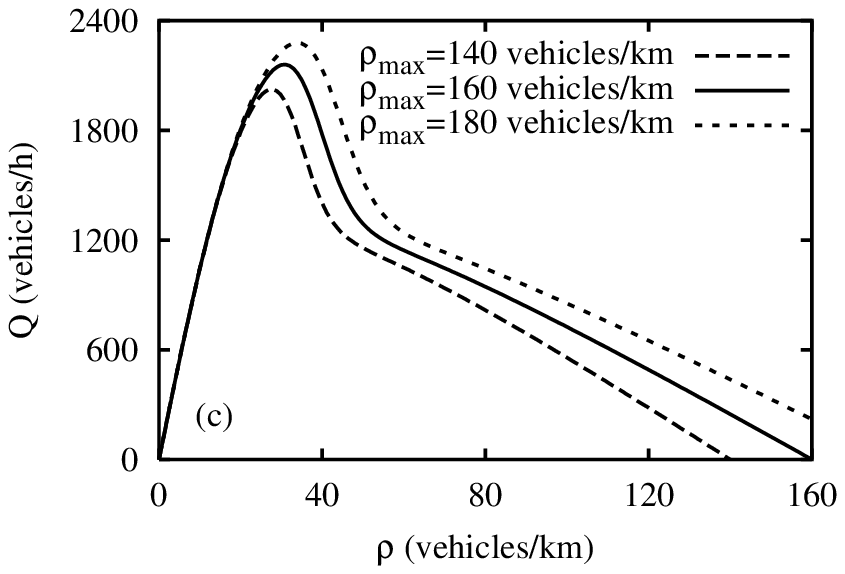} \\[-0mm]
   \includegraphics[width=90\unitlength]{\pathfigs/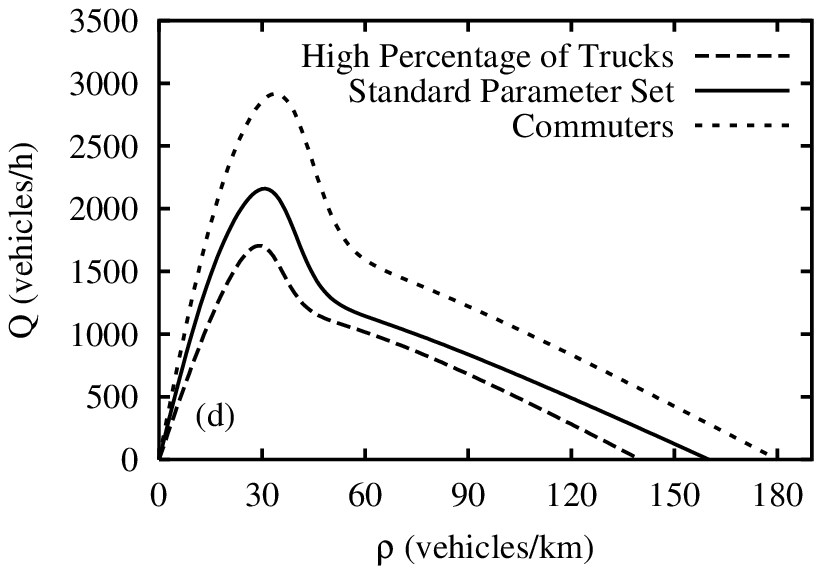} \\[-0mm]
\end{center}

\caption[]{ \label{fig_eqQ} Equilibrium flow-density relations of
the GKT model.
The diagrams (a) through (c) show the variation with
the model parameters $V_0$, $T$, and $\rho_{\rm max}$.
In each diagram, the solid lines correspond to the
standard parameter set displayed in Table 
\protect\ref{tab_par}.
Diagram (d) shows parameter combinations as one would expect
for, e.g., an increased percentage of trucks 
($V_0 = 80$\,km/h, $T= 1.8$\,s, $\rho_{\rm max} = 140$\,vehicles/km),
or a decreased percentage
($V_0 = 140$\,km/h, $T= 1.4$\,s, $\rho_{\rm max} = 180$\,vehicles/km).
}
\end{figure}

In the plots of Fig. \ref{fig_eqQ}, the resulting equilibrium flow
$Q_{\rm e} = \rho V_{\rm e}(\rho)$ is plotted as a function of the density for
various values of the model parameters $V_0$, $T$, and $\rhomax$.
The solid lines in the plots \ref{fig_eqQ}(a) - 
\ref{fig_eqQ}(d) show the result for
the standard parameter set of Table \ref{tab_par}.

\begin{figure}
\begin{center}
   \includegraphics[width=90\unitlength]{\pathfigs/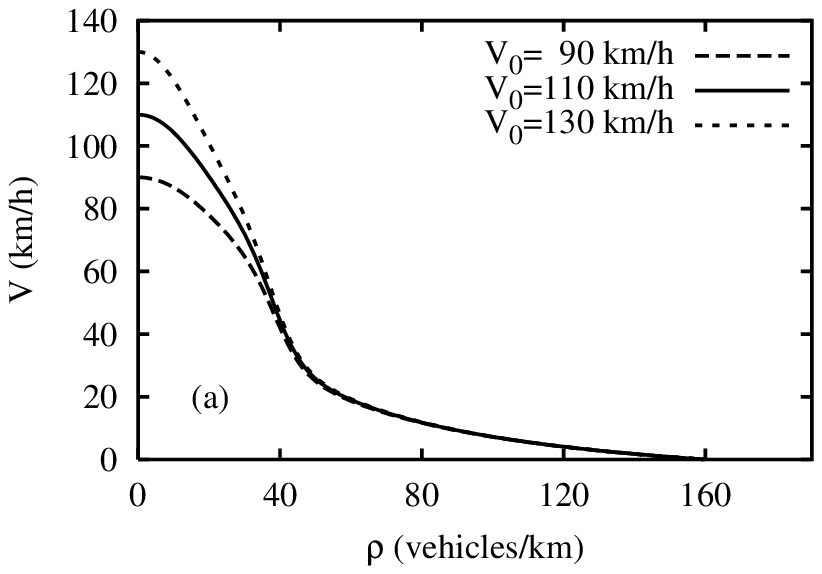} \\[-0mm]
   \includegraphics[width=90\unitlength]{\pathfigs/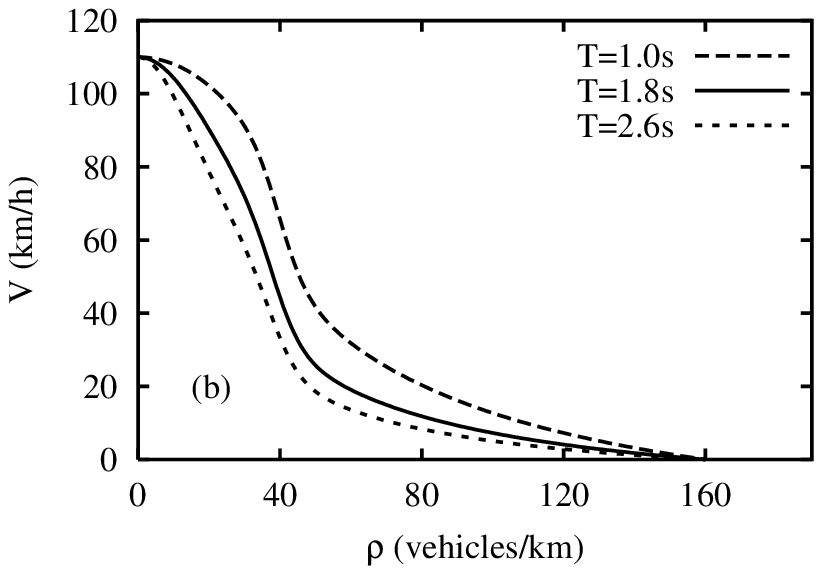}
\end{center}

\caption[]{ \label{fig_eqV} Variation of the  equilibrium 
velocity-density relation of the GKT model
with (a) the desired velocity $V_0$,
and (b) the safe time headway $T$. 
The other parameters have the standard values displayed in
Table  \protect\ref{tab_par}.
}
\end{figure}

Figures \ref{fig_eqQ}(a) and (b) illustrate 
that, according to our model,  traffic flow will increase with growing
desired velocity $V_0$ and with decreasing safe time headway $T$, as expected.
The desired velocity $V_0$ mainly influences the low-density regime, 
while $T$ is relevant for high densities.
This is also underlined by Figs. \ref{fig_eqV}(a) and \ref{fig_eqV}(b),
which show the equilibrium velocity
for the same parameters. As expected, a reduction of $V_0$
(e.g. caused by a speed limit or an uphill gradient) 
influences traffic only
at low densities. Since a reduction of $V_0$ increases the stability
of homogeneous traffic (see below), this has practical implications.
A variable speed limit, which is active only above a certain density
threshold, would increase the stability 
without reducing the capacity of the road.

The changes of traffic flow resulting from variations of $\rhomax$
are plausible as well [Fig. \ref{fig_eqQ}(c)].
Decreasing $\rhomax$ or, equivalently,
increasing the average vehicle
length (corresponding to a higher percentage of trucks) reduces
the maximum flow (capacity) of the road.
Of course, a higher proportion of trucks  also leads to a lower
desired velocity. These combined effects 
are shown in Fig. \ref{fig_eqQ}(d).

\subsection{Stability of Homogeneous Traffic
With Respect to a Localized Perturbation}
\label{stab}
%
The homogeneous
and stationary equilibrium solution \refkl{Ve}
investigated in Sec.~\ref{hom_traff} is not always stable. 
Here, we consider its stability with respect
to a localized perturbation in the
initial conditions.
Specifically,  we assume a dipole-like initial variation of the
average density $\overline{\rho}$ according to 
\begin{eqnarray}
\label{pertKK}
\rho(x,0) = \overline{\rho}
      &+& \Delta \rho \left[
          \cosh^{-2} \left(\frac{x-x_0}{w^+} \right) \right. 
                   \nonumber  \\
      &-& \left. \frac{w^+}{w^-} 
          \cosh^{-2} \left(\frac{x-x_0-\Delta x_0}{w^-} \right)
        \right] \, ,
\end{eqnarray}
as suggested in Ref. \cite{kerner-comp}.
The positive and negative peaks are positioned at 
$x_0$ and $x_0 + \Delta x_0$, respectively, with $\Delta x_0$ = 1006.25 m, 
and they have the widths
$w^+$ = 201.25 m, and 
$w^-$ = 805 m, respectively.
The amplitude $\Delta \rho$ of the perturbation 
will be varied in the simulations.
The initial flow $Q(x,0) = Q_{\rm e}( \rho(x,0))$ is assumed to be 
in local equilibrium everywhere. Our
simulations showed that the specific shape of the perturbation
is not relevant. Moreover, localized
perturbations of a different form
(e.g., a constant density plus
a Gaussian perturbation of the average velocity)
were, after a short time,  transformed to dipole-like
perturbations similar to that of Eq. \refkl{pertKK}.
The simulations were carried out with explicit
finite-difference methods \cite{num}.

\begin{figure}

\begin{center}
   \includegraphics[width=110\unitlength]{\pathfigs/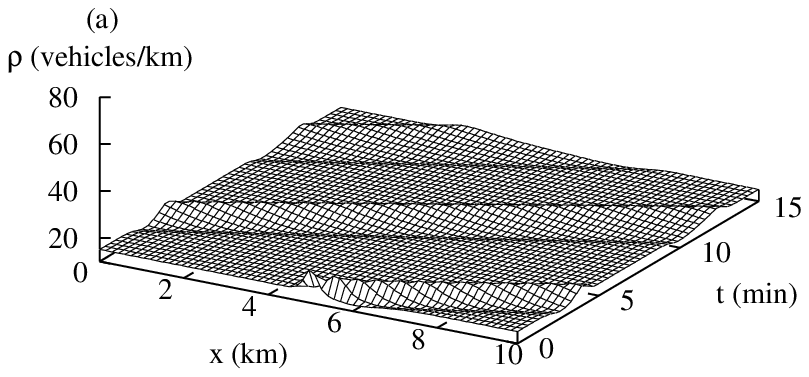} \\[-0mm]
   \includegraphics[width=110\unitlength]{\pathfigs/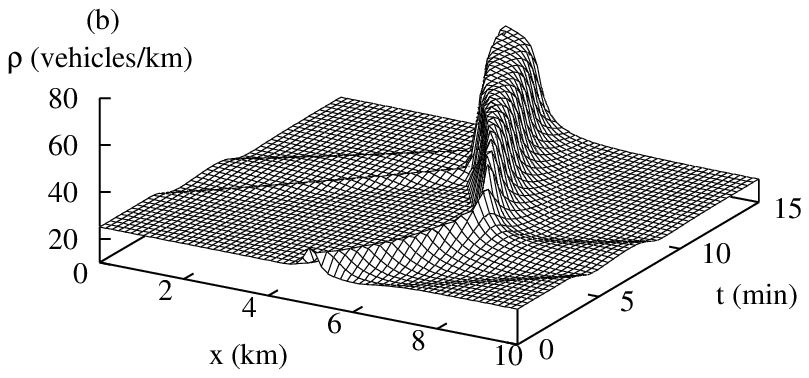} \\[-0mm]
   \includegraphics[width=110\unitlength]{\pathfigs/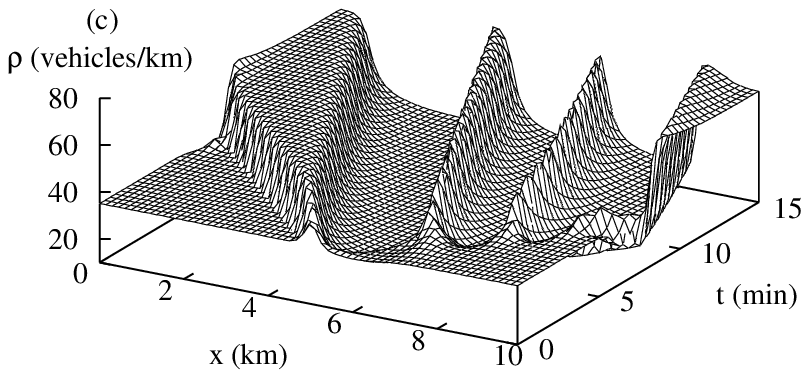} \\[-0mm]
   \includegraphics[width=110\unitlength]{\pathfigs/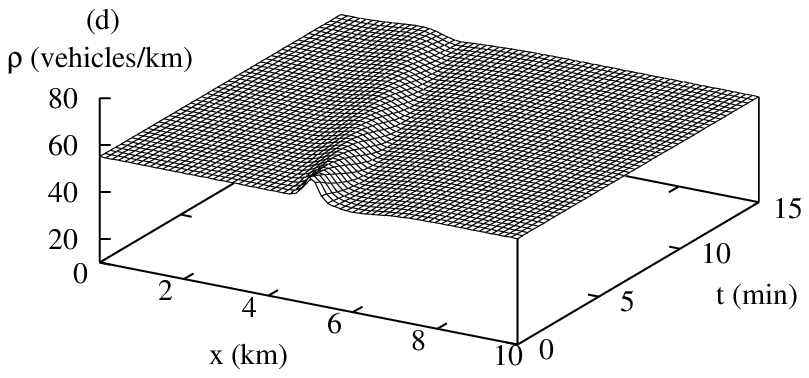} \\[-0mm]
\end{center}

\caption[]{ \label{fig_rho3D} Spatio-temporal evolution of 
traffic on a uni-directional ring of circumference 10 km,
starting with initially homogeneous traffic 
(of density $\overline{\rho}$) to which
a  localized perturbation of amplitude $\Delta \rho = 10$\, vehicles/km
is added in accordance with Eq. (\ref{pertKK}). 
(a) is for $\overline{\rho}$ = 15\,vehicles/km
(linearly stable),
(b) and (c) are for unstable traffic
($\overline{\rho}$ = 25 vehicles/km and 35 vehicles/km),
and (d) is for stable congested traffic (55 vehicles/km).
The model parameters are given by the standard set displayed in
Table \protect\ref{tab_par}.
}

\end{figure}

Figure \ref{fig_rho3D}
shows the spatio-temporal evolution of the initial perturbation
\refkl{pertKK} 
with $\Delta\rho$  = 10 vehicles/km
for various  initial densities $\overline{\rho}$.
In Fig. \ref{fig_rho3D}(a), it is
shown that the perturbation dissipates, if the traffic
density is sufficiently low.
When increasing the initial density, perturbations
eventually lead to
instabilities. Depending on the density, a single density cluster 
[cf. Fig. \ref{fig_rho3D}(b)], 
or a cascade of traffic jams (i.e., stop-and-go traffic) [Fig. 
\ref{fig_rho3D}(c)] is triggered.
If one further increases the density, we observed dipole-like
structures similarly to those described in Ref.
\cite{kerner-dipole}. Finally, for densities above 50-55\,vehicles/km,
one reaches
again a stable regime  [Fig. \ref{fig_rho3D}(d)].

The underlying instability mechanism is intuitive \cite{kk-94}.
In the perturbation \refkl{pertKK}, the positive density peak
(with a lower velocity than in the homogeneous regions)
is behind the negative peak (with higher velocity).
This means, drivers in the region upstream of the perturbation will
approach the positive peak. If the traffic density is sufficiently low, 
the vehicles  can overtake by lane changing without braking,
as soon as they meet the tail of the perturbation. 
In addition, the perturbation dissolves by the 
above described effect of kinematic dispersion. 
As a result, homogeneous traffic is stable at low densities.
For higher densities, however, a higher percentage of drivers
approaching the density peak must brake, thereby locally increasing the
density. The increased density, in turn, gives a positive
feedback for further braking reactions.
This feedback cycle continues until the resulting velocity is so low
that the acceleration term compensates the braking effects.
This defines the jammed density
$\rho = \rho_{\rm jam}$ with nonzero flow 
$Q = Q_{\rm e}(\rho_{\rm jam})$.
Furthermore, it makes plausible
that homogeneous
traffic of density $\overline{\rho}\ge\rho_{\rm jam}$ is stable again.

\begin{figure}
\begin{center}
   \includegraphics[width=90\unitlength]{\pathfigs/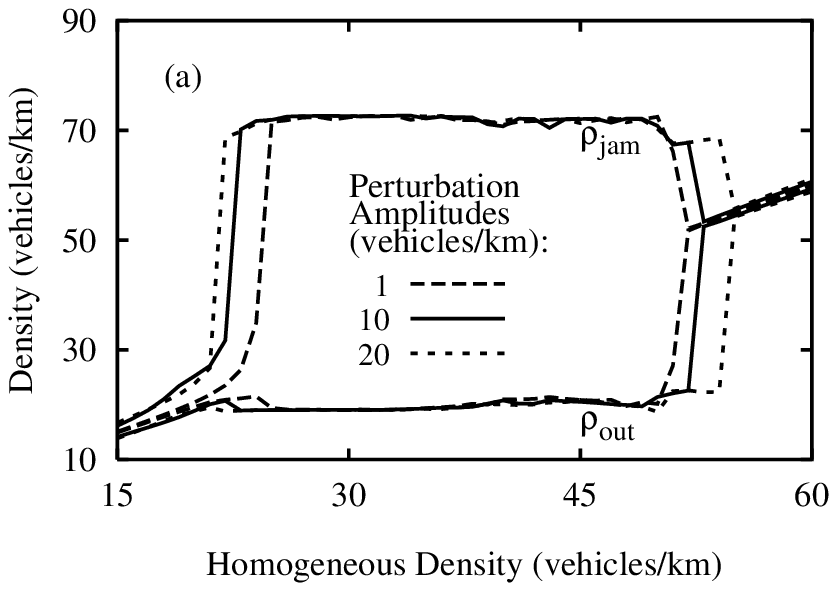}  \\[-0mm]
   \includegraphics[width=90\unitlength]{\pathfigs/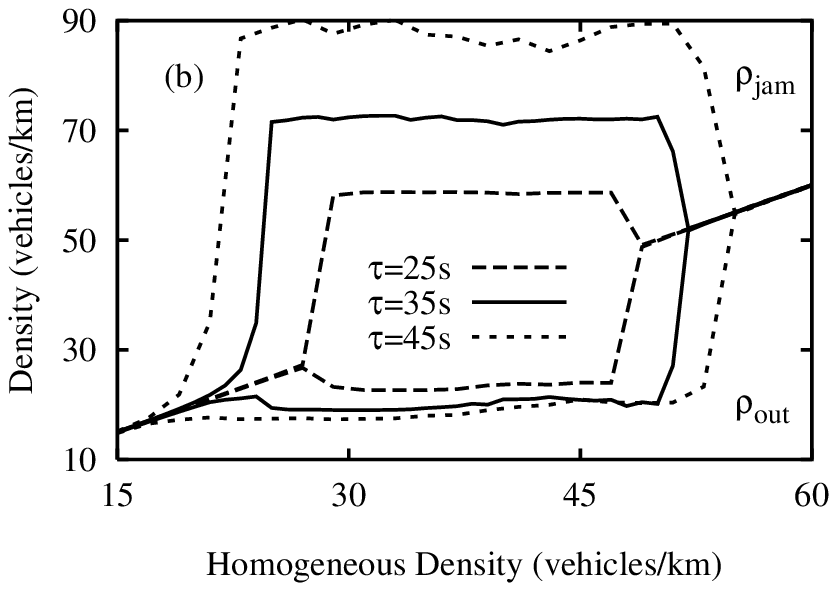} 
\end{center}

\caption[]{ \label{fig_phase} Stability diagram 
for perturbations of homogeneous traffic of the form
\protect\refkl{pertKK} in
a ring of circumference 10 km.
Both diagrams show
the maximum and minimum densities $\rho_{\rm jam}$
and  $\rho_{\rm out}$ as a function of the average density
$\overline{\rho}$,
measured after a dynamical equilibrium was reached. 
The unstable traffic regime corresponds to the density range where 
the jam amplitude
$(\rho_{\rm jam} - \rho_{\rm out})$ is large (rectangle-like
shaped regions).
Diagram (a) shows the dependence of the stability diagram on the 
perturbation amplitude $\Delta \rho$.
One can clearly see two density ranges  
$[\rho_{\rm c1},\rho_{\rm c2}]$ with
$\rho_{\rm c1} =$\, 21 
vehicles/km and $\rho_{\rm c2} = $\, 24 
vehicles/km, and
$[\rho_{\rm c3},\rho_{\rm c4}]$ with
$\rho_{\rm c3} =$\, 51 
vehicles/km and $\rho_{\rm c4} = $\, 55 
vehicles/km, where  traffic is non-linearly stable, i.e.,
stable for small perturbations, but unstable for large
perturbations. In the range
$[\rho_{\rm c2},\rho_{\rm c3}]$, 
homogeneous traffic is unstable for arbitrary
perturbation amplitudes.
Diagram (b) shows the stability diagram for various relaxation times
$\tau$ and a perturbation
amplitude of $\Delta\rho = 1$\,vehicle/km.
}

\end{figure}

An important criterium for realistic traffic models is, that 
the transition to these inhomogeneous states 
should be hysteretic, corresponding to a
first-order phase transition  \cite{kk-94}.
This implies that, at least in some parameter range, the 
response of the system to localized perturbation depends
on the perturbation amplitude  (bistability).
In terms of real traffic, there are situations
where traffic flow 
is metastable with respect to small perturbations,
but it breaks down if the perturbations are sufficiently large.
Figure \ref{fig_phase} shows that we found  two ranges of densities,
$\overline{\rho}\in [\rho_{\rm c1}, \rho_{\rm c2}]$ and
$\overline{\rho}\in [\rho_{\rm c3}, \rho_{\rm c4}]$,
where the transition is bistable. 
For large  perturbation amplitudes, the system develops
to a localized-cluster state 
if $\overline{\rho}\in [\rho_{\rm c1}, \rho_{\rm c2}]$,
or to a dipole-like state
if $\overline{\rho}\in [\rho_{\rm c3}, \rho_{\rm c4}]$,
while  it relaxes back to the equilibrium state
for smaller perturbations.
Between these ranges, there is a region 
$\overline{\rho}\in [\rho_{\rm c2}, \rho_{\rm c3}]$, where
homogeneous traffic is linearly unstable, giving rise to cascades
of density clusters (``stop-and-go waves'').
For $\rho<\rho_{\rm c1}$ and $\rho>\rho_{\rm c4}$, traffic is stable
for arbitrary perturbations.

\begin{figure}
\begin{center}
   \includegraphics[width=90\unitlength]{\pathfigs/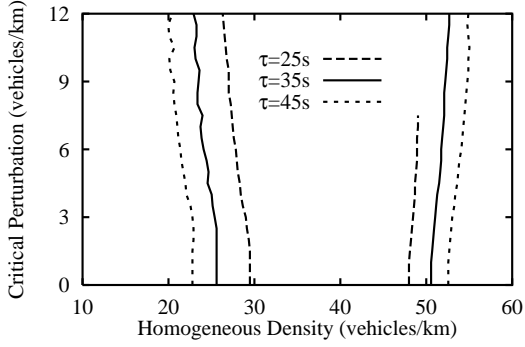} 
\end{center}

\caption[]{ \label{fig_hyst} 
Critical perturbation
amplitudes $\Delta\rho_{\rm c}$ of localized perturbations
in the metastable (non-linearly unstable) density regime for various
values of $\tau$.
The critical amplitude is the minimal amplitude that can cause
traffic jam formation. Smaller amplitudes are damped out in the course of time.
Between the two metastable density regions arbitrarily small perturbation
amplitudes will cause the formation of traffic jams. At low and high
densities, inhomogeneities of traffic flow tend to disappear.
}
\end{figure}

In Fig. \ref{fig_phase}(a), we plot the 
minimum and maximum densities
$\rho_{\rm min}$ and $\rho_{\rm jam}$ that resulted 
from simulations of traffic on a circular road
after  a dynamic steady state has 
been reached.
A large difference $(\rho_{\rm jam}-\rho_{\rm min})$
corresponds to  localized-cluster or stop-and-go states, while 
$(\rho_{\rm jam}-\rho_{\rm min}) \approx 0$
is the signature of the homogeneous state.
At a perturbation amplitude of one vehicle per km (dashed lines),
the perturbation can be
considered as being linear,  while at 
$\Delta\rho=$ 20 vehicles/km (dotted lines), the perturbation amplitude 
reaches the order of the homogeneous densities 
defining the maximum perturbation. So, we can determine
from the plot the four critical densities.
Notice that the densities $\rho_{\rm jam}$ and $\rho_{\rm min}$ 
of the developed non-linear state  depend neither  on the initial
density nor on the amplitude of the perturbation.
Figure \ref{fig_hyst} shows the basins of attraction 
of the two traffic states
in the phase space spanned by $\overline{\rho}$ and $\Delta\rho$.
For values of $(\overline{\rho},\Delta\rho)$ corresponding to
points inside the two curves, the final steady-state consists of
density clusters or stop-and-go waves.
Otherwise, the final state is that of homogeneous equilibrium.

Now, we will show that the stability of the model is determined mainly by
the relaxation time $\tau$ and by the anticipation factor $\gamma$. 
Since the flow-density relation does not depend on these parameters,
this means that one can calibrate the stability and the flow rates 
independently.
Figures \ref{fig_phase}(b) and
\ref{fig_hyst} show that an increased value of
$\tau$ leads both to an increased range of instability, and
to increased amplitudes 
$(\rho_{\rm jam} - \rho_{\rm min})$ of the non-linear state.
If $\tau$ exceeds values of about 60 s, the density exceeds 
the maximum density $\rhomax$ in the course of the simulations, 
which is a
signature of accidents.
For $\tau \le $ 12 s, the unstable and metastable regions vanish altogether,
and the system is globally stable for all densities and all
perturbation amplitudes.
For a fixed value $\gamma=1.2$ but otherwise arbitrary  model parameters,
global stability is reached, if the dimensionless parameter
$V'_0=\rhomax V_0\tau$ of the scaled equations
\refkl{eqrhoscal} and \refkl{eqVscal} satisfies 
$V'_0\le V'_{\rm c} = 59$.
Further simulations showed that the  stability of 
traffic described by the GKT model 
increases also with  increasing  $\gamma$.
Both the critical relaxation time where accidents occur, and the
critical value $V'_{\rm c}$ for global stability increase.
This is plausible since  $\gamma$  desribes the 
anticipation of  future velocity changes.

It turned out that the qualitative difference
between  the non-linear state of Fig. \ref{fig_rho3D}(b), where only one
cluster is seen, and that of Fig.
\ref{fig_rho3D}(c),
where a cascade of stop-and-go waves appeared,
can be understood by the linearly unstable and metastable
density ranges discussed above.
In both cases, there is a region in the wake of the first cluster, 
where the density is lower, and the 
velocity is higher than in the homogeneous region downstream.
This gives rise to a transition layer between these regions
serving  as a 
small perturbation \cite{kk-94}. In Fig. \ref{fig_rho3D}(b), the system is 
metastable,
and this perturbation is too small to induce a new density cluster.
In  Fig. \ref{fig_rho3D}(c), the system is linearly unstable
and the transition layer triggers a new cluster.
This cluster, in turn, gives rise
to a new transition layer eventually leading
to a cascade of stop-and-go waves.

\begin{figure}
\begin{center}
   \includegraphics[width=90\unitlength]{\pathfigs/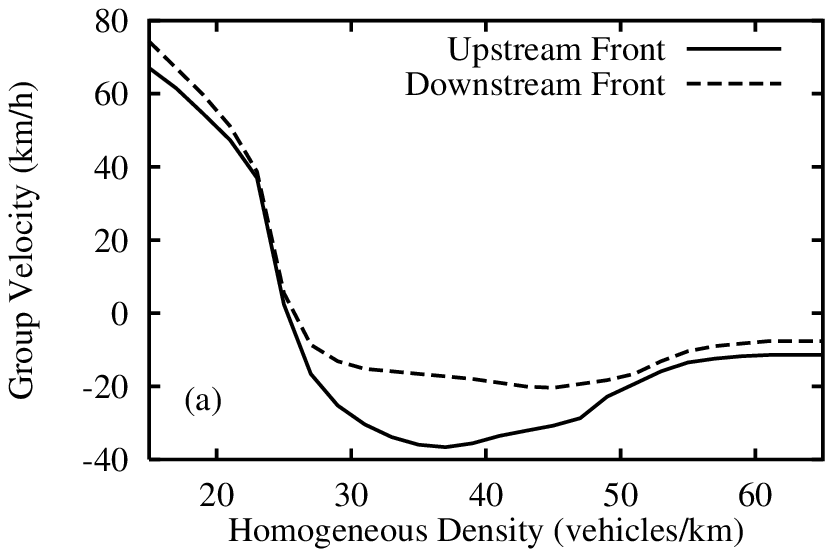} 
   \includegraphics[width=90\unitlength]{\pathfigs/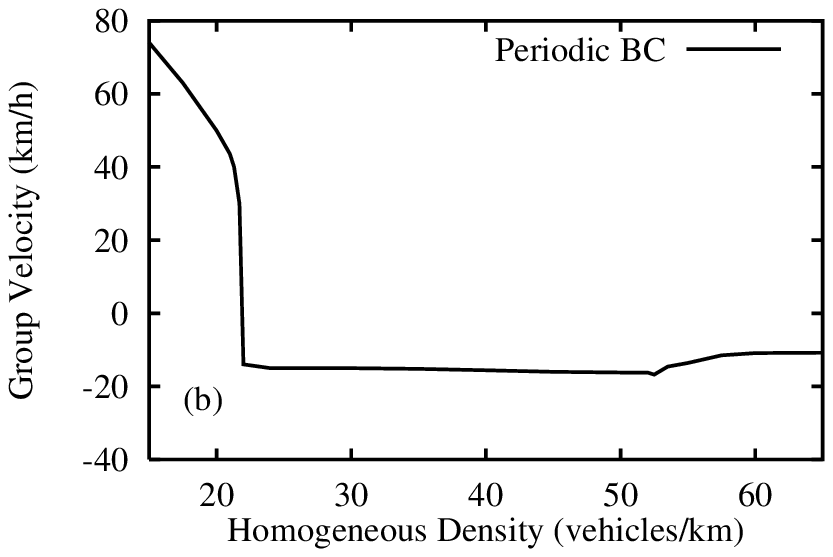} 
\end{center}

\caption[]{ \label{fig_vg} (a) Group velocity of the upstream front (solid)
and downstream front (dashed)
of the first stop-and-go wave displayed in Fig. \protect\ref{fig_rho3D}.
The propagation of the fronts is calculated during a period,
where the density of traffic jams does not grow anymore ($t>4$\,min),
but the dynamics does not yet depend on the boundary conditions
($t< 15$ min). Notice that the propagation velocity at low densities is
positive, but slower than the average vehicle velocity. In the 
instability region, the negative propagation 
velocity of the downstream front depends
only weakly on the initial density, and its magnitude is well compatible
with empirical data. The larger propagation velocity of the upstream
front comes from the growing jam length. (b) Group velocity
of the upstream and downstream fronts on a circular road.
Here, the length of traffic jams stabilizes after some time, 
and their upstream fronts
move with the same velocities as their downstream fronts. 
The magnitude of the initial perturbation ($\Delta\rho$ = 10
vehicles/km)
has been chosen larger than in diagram 
(a) ($\Delta\rho$ =1 vehicle/km), resulting 
in a larger region of negative group velocities.
}

\end{figure}

Another remarkable feature of Fig. \ref{fig_rho3D}(c) is
that the width  of the first (upstream) stop-and go wave
is growing while the other waves remain narrow.
The  width of the cluster in Fig. \ref{fig_rho3D}(b) increases as
well, but slower. This can be explained by
observing that the growth rate of the width is given by the difference
of the  group velocities
$v_{\rm g,up} = (Q_{\rm jam} - Q_{\rm in}) / (\rho_{\rm jam} - \rho_{\rm in})$
and
$v_{\rm g} = (Q_{\rm out} - Q_{\rm jam}) / (\rho_{\rm out} - \rho_{\rm jam})$
of the upstream and downstream fronts 
(Fig. \ref{fig_vg}).
For the density $\overline{\rho}$ = 25 vehicles/km
corresponding to Fig. \ref{fig_rho3D}(b), the difference is
small, while  for $\overline{\rho}$ = 35 vehicles/km
[Fig. \ref{fig_rho3D}(c)], it is large.
In all subsequent clusters of Fig. \ref{fig_rho3D}(c), 
we have $Q_{\rm in} = Q_{\rm out}$ and
$\rho_{\rm in} = \rho_{\rm out}$, so they do not grow.

\begin{figure}
\begin{center}
  \includegraphics[width=90\unitlength]{\pathfigs/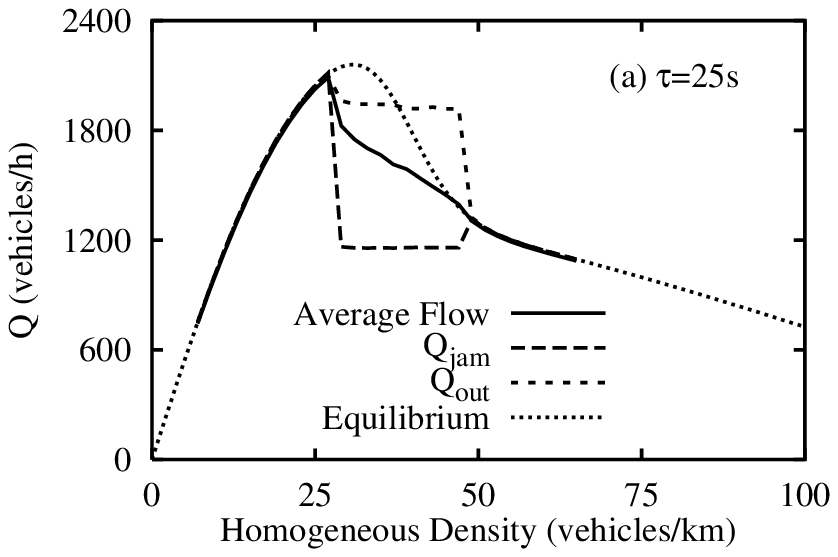} \\[0mm]
  \includegraphics[width=90\unitlength]{\pathfigs/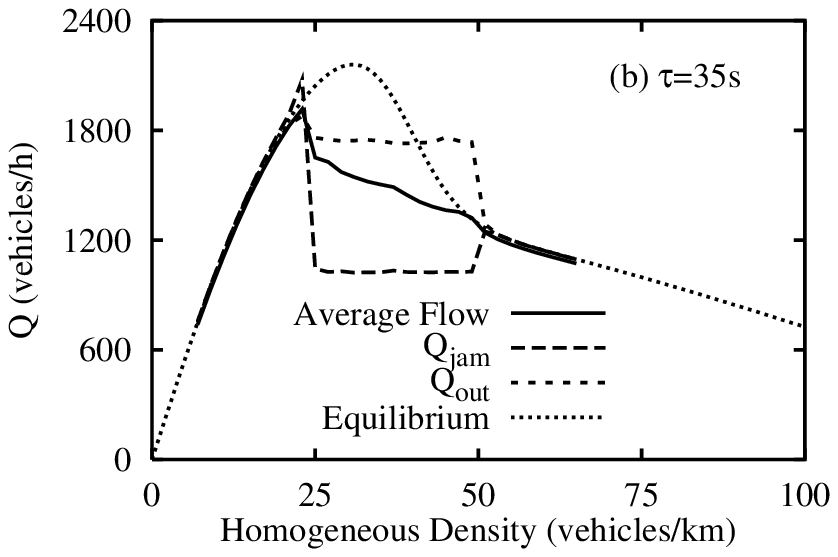}     \\[0mm]
  \includegraphics[width=90\unitlength]{\pathfigs/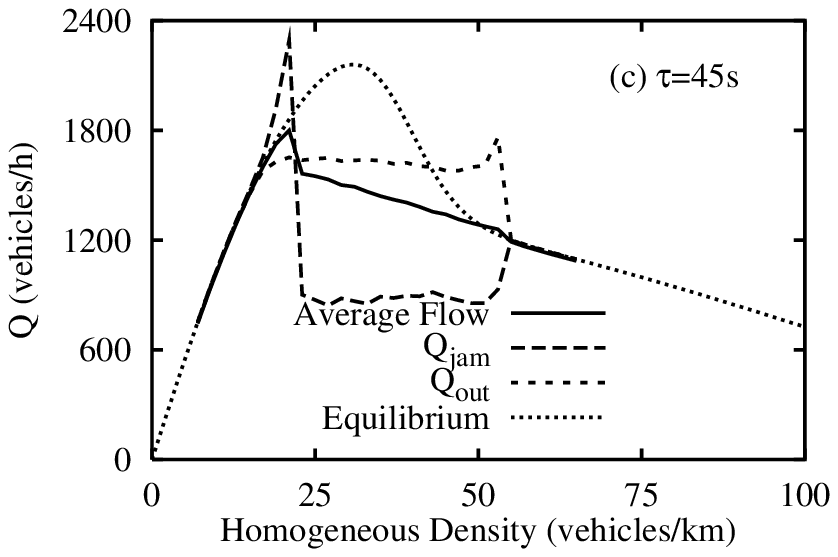} 
\end{center}

\caption[]{ \label{fig_avgQ} Characteristic flows in 
the fully developed stop-and-go
traffic corresponding to Fig. \protect\ref{fig_phase}(b), which
results on a circular road from locally perturbed traffic
of average density $\overline{\rho}$. 
Depicted are, as a function of $\overline{\rho}$,
the flows $Q_{\rm jam}$ in the jammed regions,
the outflows $Q_{\rm out}$ from jams,
and the average flows. For comparison, the
equilibrium flow $Q_{\rm e} = \rho V_{\rm e}$ with $V_{\rm e}$ from
Eq. \protect\refkl{Ve} is also shown.
Notice that, in the unstable range, the average {\it dynamic} flow is
lower than the equilibrium flow.
}

\end{figure}

It is required for
any realistic traffic model,
that the density $\rho_{\rm jam}$ inside localized clusters
and the traffic in the region downstream of it is
independent of the inflow \cite{kk-94}.
This implies that the group velocity $v_{\rm g}$
should be constant.
The upper curves in Fig. \ref{fig_phase} for the jam density,
the  dashed line of Fig. \ref{fig_vg} for $v_{\rm g}$, and
the dashed line of Fig. \ref{fig_avgQ} for $Q_{\rm out}$ show, that the
GKT model essentially fulfills this requirement.

\begin{figure}
\begin{center}
   \includegraphics[width=90\unitlength]{\pathfigs/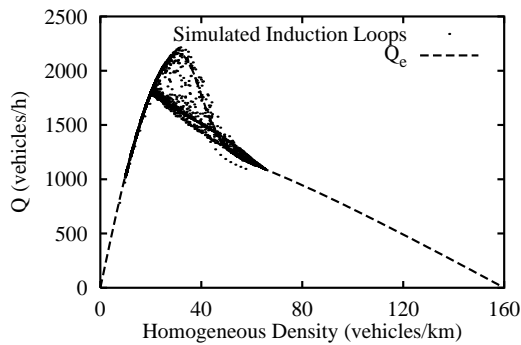} \\[0mm]
\end{center}

\caption[]{ \label{fig_Qtheodots} 
Simulated flow-density relation as superposition
of data obtained at five equally distributed ``detectors'' on
a circular road of length 10\,km (dots). In order to take into account 
the variation of traffic densities in the course of the day,
the average density $\overline{\rho}$ was 
varied in the range [10\,vehicles/km, 70\,vehicles/km] in steps of 2 veh/km.
The total duration of simulations for the homogeneous density
$\overline{\rho}$ was $70\,\mbox{min} [1-\overline{\rho}/
(70\,\mbox{vehicles/km})]$. Note that, in the unstable regime,
the dynamical flow-density values tend
to lie below the equilibrium relation (dashed line).
}

\end{figure}

Another frequently discussed problem is, 
to which extent equilibrium flow-density
relations of traffic models can be calibrated by non-equilibrium
empirical data.
In Fig. \ref{fig_Qtheodots} we simulate the measurement of traffic
data  by recording 
the velocity and density at several fixed locations 
(``measuring cross-sections'')
over a certain period of time. To incorporate the distribution 
of  densities encountered in real traffic, we run simulations
for various densities. In each run, the simulation time was proportional to
the occurence probability of 
traffic densities in the course of the day, for which we 
assumed a linear decrease from a maximum value
at $\overline{\rho} = 0$ to
zero for $\overline{\rho}$ = 70 vehicles/km.
The dots in Figure \ref{fig_Qtheodots} show the flows and densities
``measured'' at the cross-sections for all simulation runs put together.
In addition, the solid line in
Fig. \ref{fig_avgQ} shows the dynamic average of the flow
for runs at a given density. Both figures show
that, in the case of unstable traffic,
the flow of the ``measured'' dynamic flow-density relation
(broken line) is considerably lower than the
equilibrium flow. 
From this it follows that, in regions of unstable traffic,
one cannot calibrate equilibrium
flows of traffic models to empirical flow-density relations. 

\subsection{Fronts Between Congested and Free Traffic}
\label{fronts}

\begin{figure}
\centering\epsfig{file=\pathfigs/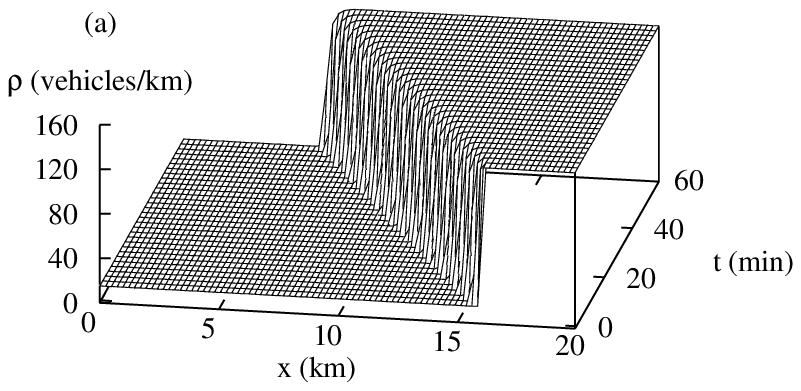,width=110\unitlength}
\centering\epsfig{file=\pathfigs/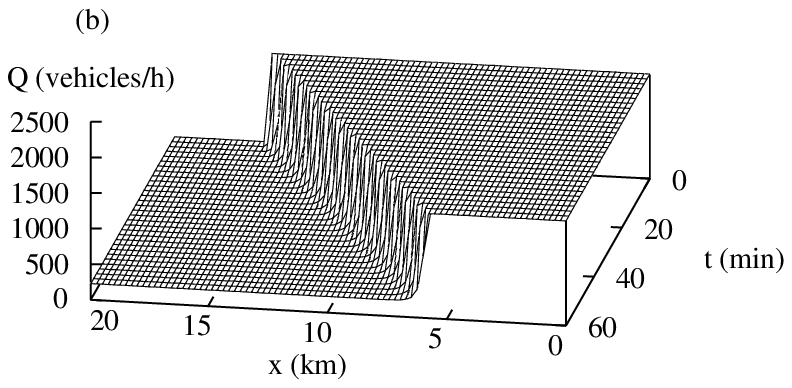,width=110\unitlength}
\caption[]{
Simulation of an upstream front with initial densities
of $\rho_{1}=15$ vehicles/km
and $\rho_{2}=140$ vehicles/km.
Shown is the spatio-temporal evolution of (a) the density, and (b) the flow.
In (b), the direction of the space and time axes is reversed for
illustrative reasons.
}
\label{fig_upfront}
\end{figure}

The realistic description of shock fronts in traffic is a particularly
difficult problem, as pointed out in Ref.~\cite{Dag}.
Therefore, in this section we will investigate how fronts between two
different states of traffic, especially between free and congested
traffic, evolve in the GKT model.
We model such fronts by initial conditions containing
discontinuities in 
the fields $\rho$ and $V$.
In particular, we consider shocks at $x_{\rm shock}$ between two homogeneous
regions with densities $\rho_{1}$ and
$\rho_{2}$ at the left 
(upstream) and right (downstream) sides.  
Initial conditions with $\rho_{1} > \rho_{2}$ include 
situations of dissolving jams.
The extreme case $\rho_{1}=\rho_{\rm max}$ corresponds 
to vehicles starting from 
standstill after a road blockage is removed.
Initial conditions with $\rho_{1} < \rho_{2}$ include  situations,
where free traffic flow meets a queue of nearly standing vehicles.
The simulations were carried out with the standard
parameter set of Table \ref{tab_par}.
At the upstream boundary ($x=0$\,km), we chose the fixed (Dirichlet) boundary
conditions $\rho(0,t)=\rho_{1}$ and $Q(0,t)=Q_{e}(\rho_{1})$ to model a 
constant inflow consistent with the initial conditions. 
At the downstream boundary ($x_{\rm max}$ = 40 km), we simulated an unperturbed 
outflow
by the homogeneous von Neumann conditions $\partial_{x} \rho(x_{\rm max},t)$ = 
$\partial_{x} Q(x_{\rm max},t)$ = 0.
Figure \ref{fig_upfront}(a) and \ref{fig_upfront}(b) show the spatio-temporal 
development of the density and the flow for  $\rho_{1} < \rho_{2}$ 
(upstream jam-fronts).


One can see that the shape of the backwards moving front does not change in time
[cf. Fig.\ref{fig_upfront}(a)] and that there is no region of negative velocity
[Fig.\ref{fig_upfront}(b)].
This is achieved by the non-local interaction term in the velocity equation, while
a viscosity term ($\sim \partial^2 V/ \partial x^{2}$) would make the front
smoother with increasing time.

\begin{figure}
\centering\epsfig{file=\pathfigs/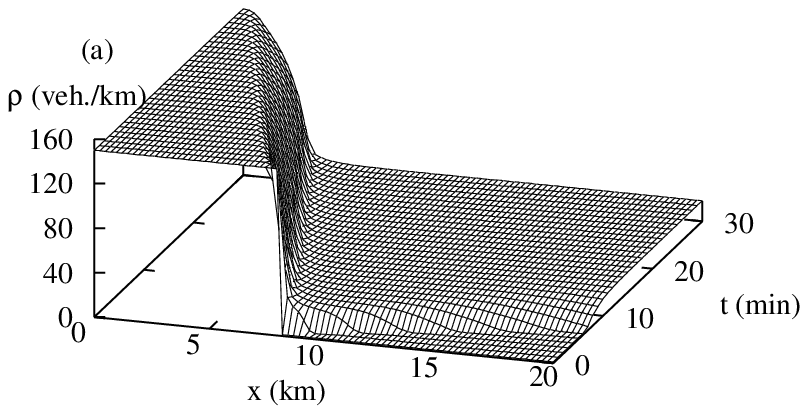,width=110\unitlength}
\centering\epsfig{file=\pathfigs/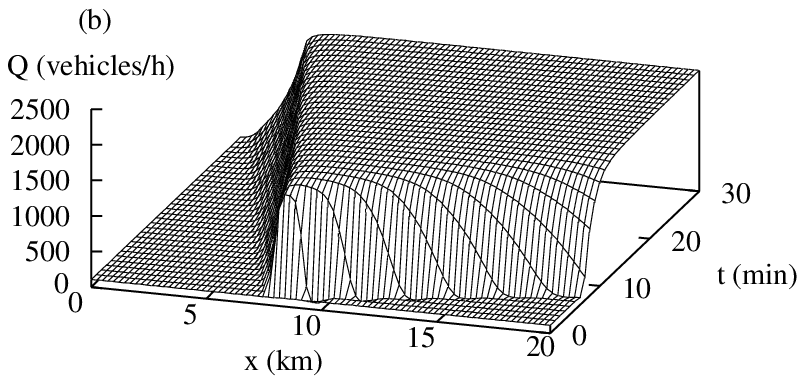,width=110\unitlength}
\caption[]{Simulations of downstream-fronts 
with $\rho_{\rm jam}$ = 140 vehicles/km.
Shown is the spatio-temporal development of (a) the density, and (b)
the flow. Free boundary conditions were used on both sides.}
\label{fig_downfront}
\end{figure}

Figures \ref{fig_downfront}(a) and \ref{fig_downfront}(b) 
show the spatio-temporal evolution 
of the density and flow for downstream fronts
with $\rho_{1} \gg \rho_{2}$. This corresponds to a dissolving jam 
(for example, after an accident has been cleared) with an outflow
to a nearly empty road section. Due to the kinematic dispersion term
$(1/\rho)\partial(\rho \theta) /\partial x$ in the velocity equation,
the forward moving front is somewhat smoothed
in the course of time.
The finite velocity variance implies that, after some time, the faster cars are 
found in a wider distance from the jam front.

\begin{figure}


\centering\epsfig{file=\pathfigs/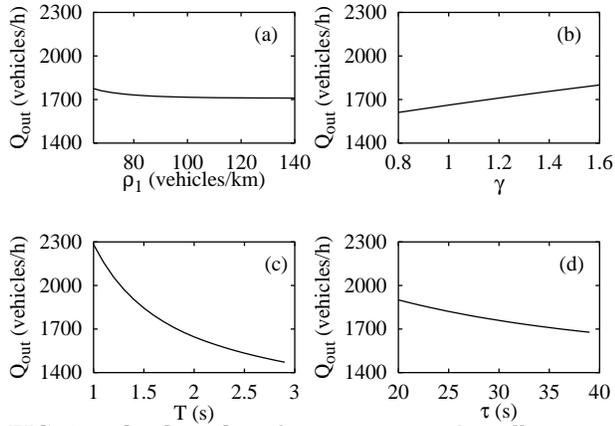,width=95\unitlength}

\caption[]{Outflow $Q_{\rm out}$ from a congested traffic state of
density $\rho_{\rm 1}$ in dependence of 
(a) $\rho_{\rm 1}$,
(b) $\gamma$,
(c) $T$,
and (d) $\tau$.
The simulated traffic situation is that of Fig. \protect\ref{fig_downfront}.
The outflows were determined after a transient time of
30 minutes.
}
\label{fig:qout1}
\end{figure}

As a remarkable fact it should be mentioned that, although the 
equilibrium point of the outflow $(\rho_{\rm out}$,
$Q_{\rm e}(\rho_{\rm out}))$ 
is the  result of a dynamic process, the outflow $Q_{\rm out}$ is nearly
constant over a wide range of $\rho_{1}$ [cf. Fig.\ref{fig:qout1}(a)]. 
This agrees with empirical observations where it has been found
that  the
outflow of very different forms of congested traffic
(including the dissolution of queued city traffic after a traffic light
turns green)
is nearly a ``constant of traffic''
\cite{kerner-chania,Ref}.

This generalizes the above mentioned requirement of a constant outflow from
clusters (whose jam density is determined by the dynamics) to forms of
congestion that are a result of the initial and boundary conditions.

However, the outflow $Q_{\rm out}$ varies with the model parameters.
It increases with growing $\gamma$ [cf. Fig.\ref{fig:qout1}(b)],
because  an acceleration tendency
is already recognized  in a larger 
distance from the jam front.
The outflow decreases drastically with increasing $T$ 
[cf. Fig.\ref{fig:qout1}(c)],
because $T$ determines the time headway between two 
following vehicles or, equivalently, the inverse of
the flow.
Furthermore, the outflow is diminished with
increased $\tau$ [cf. Fig\ref{fig:qout1}(d)], because  increased
values of $\tau$ correspond to lower accelerations and thus to more
inert vehicles.

\subsection{Method of Parameter Calibration}
\label{results_diss}

While the model parameters $V_0$, $T$, and $\rhomax$ influence the
equilibrium flow-density (or velocity-density) relation
(Figs. \ref{fig_eqQ} and \ref{fig_eqV}), 
the parameters $\tau$ (Fig. \ref{fig_phase}) and
$\gamma$ influence only the stability behavior.
This enables an effective calibration of the GKT model to
concrete traffic situations.

First, $V_0$ is determined as the low-density limit
of the experimental velocity-density relation (which, in this limit,
is also the equilibrium relation), and $\rhomax$ is determined
by  the average length of the vehicles and assuming
a reasonable bumper-to bumper distance of, e.g., 
1.5\,m in standing traffic.
Then, $T$ is calibrated by the observed maximum flows, by
the outflows from stop-and-go waves, or by the flow resulting
from standing traffic after a red light turns green or
an obstacle is removed \cite{kerner-chania,Ref}.
Afterwards, one calibrates $\tau$ and $\gamma$
by the observed stability behavior 
(Fig. \ref{fig_phase}) and by the shape
and width of the downstream and upstream fronts 
connecting free and congested states
(Figs. \ref{fig_upfront} and \ref{fig_downfront}).

Since $\tau$ and $\gamma$ weakly influence the flows
[Figs. \ref{fig:qout1}(d) and \ref{fig:qout1}(b)], the calibration of
$T$, $\tau$, and $\gamma$ is repeated recursively until
convergence is obtained.

Applying this procedure to the single-vehicle data of the Dutch
motorway A9 leads to the parameter set shown in
Table \ref{tab_par}.
Notice that, because of their immediate intuitive meaning,
the plausible range of values for the model parameters
is rather restricted. One can argue that
reasonable time headways are in the range 
$T \in $ [1.0s, 2.5s], and that initial accelerations $a_{\rm max} = V_0/\tau$
are in the range $a_{\rm max} \in $ [1 m/s$^2$, 4 m/s$^2$]
corresponding (e.g. for $V_0$ = 144 km/h) to $\tau \in$ [10\,s, 40\,s].
Finally, the minimum anticipation of traffic is to the car in front,
implying $\gamma \ge 1$.

\section{Summary and Conclusions} 
\label{conclusion} 

We have proposed a macroscopic gas-kinetic-based traffic model
(GKT model) that was derived from a microscopic model of vehicle dynamics.
The assumed fluctuations of vehicle acceleration implied a 
velocity distribution of finite variance
which is governed by a kinetic equation related to that used in kinetic
gas theory.
In the resulting GKT equations, the velocity variance enters both
the braking interactions and the smoothing effect of the ``kinematic
dispersion''.  

In contrast to gas-kinetic-based models proposed earlier
\cite{helb-physica,helb-gas98}, we could 
now derive the correlation factor reflecting the
increased interaction rate of vehicles at high densities
by simple and plausible arguments.
Furthermore, we replaced the dynamic variance
by a ``constitutive relation''
obtained from  single-vehicle data,
thereby considerably simplifying the description.
The resulting relation of the flow as a function of the density
agreed well with empirical data.
Since the form of this function is determined by 
the constitutive relation, this supports the assumption
of the model that braking interactions
are mainly caused by a finite velocity  variance.

Because of its derivation from physical assumptions,
all model parameters of the GKT model have an intuitive
meaning and can be either directly measured or calibrated to real
traffic data.  It is straighforward
to model the effects of, e.g., speed limits, varying
road conditions (e.g., gradients), or different driving behavior.
Moreover, we proposed a systematic calibration procedure.

In contrast to other macroscopic traffic models, the GKT model is
non-local. While the non-locality has similar smoothing effects as
a diffusion or viscosity term, it leads to a more favourable 
numerical stability behavior. 
The model belongs to the class of effective one-lane models, so
multi-lane aspects are treated only in a global way.
Nevertheless, it is straightforward
to  include ramps \cite{helb-MT-sync}, or lane blockages
(e.g., due to road works or accidents).

Investigations of  instabilities arising from a localized perturbation
of homogeneous traffic showed qualitatively the same scenario
as has been found for the model of Kerner and Konh{\"a}user
\cite{kk-94,kerner-dipole}.
For sufficiently high values of $\tau$, homogeneous
traffic at intermediate densities becomes unstable  
with respect to perturbations, leading to non-linear states like 
localized clusters, stop-and-go waves, or dipole-like structures,
while homogeneous traffic flow is stable for high and low densities.
The GKT model satisfies two requirements which should hold for any
realistic model \cite{kk-94}. First,
the transition to localized clusters is hysteretic. Second, the outflow
region of the clusters is nearly independent of the homogeneous density.
In contrast to most other models, however, the traffic flow 
within the clusters is finite and can be influenced by the model
parameters $\tau$ and $\gamma$.
This enables the simulation of 
``synchronized'' traffic states \cite{helb-MT-sync}
that turned out to be the most frequent
form of congested traffic \cite{kerner-sync}.

Only for high densities and for $\gamma=1$, the interaction term
of the  microscopic
equations \refkl{vmicstoch} underlying the GKT model 
can be written in terms of a simple 
car-following model.
For this case,
we performed simulations with a smoothed version of 
Eq. \refkl{vmicstoch} (containing no abrupt
velocity changes in the interaction term) and
found a nearly quantitative agreement
in describing collective states like clusters or stop-and go waves. 

The high numerical stability of the GKT model
also allowed to treat realistic boundary conditions
(instead of the periodic boundary conditions used in most previous
publications), and to simulate
discontinuous fronts between homogeneous low-density
and high-density states. 
Such fronts correspond to the formation or
dissolution of jammed traffic that is
caused by initial or boundary conditions rather
than by dynamic instabilities.

Remarkably, the outflow $Q_{\rm out}$ from  jammed regions
was nearly the same as the outflow from  localized clusters,
regardless of the density of the jams, including even
standstill traffic (related, e.g., to the dissolution of
a queue behind a traffic light turning green).
The observation that the outflow from arbitrary kinds of congested
traffic is  a ``universal'' constant
of traffic dynamics, and also its numerical value
of about 1800 vehicles per hour, agrees well with 
observations of real traffic \cite{kerner-chania,Ref}.
Besides the constant group velocity of 
localized clusters, this universal outflow 
can be considered as an additional requirement for
realistic traffic models.


\subsection*{Acknowledgments}
The authors want to thank for financial support by the BMBF (research
project SANDY, grant No.~13N7092) and by the DFG (Heisenberg scholarship
He 2789/1-1). They are also grateful to Henk Taale and 
the Dutch {\it Ministry of Transport,
Public Works and Water Management} for supplying the freeway data.



\begin{references}
\bibitem{note-gamma}
Notice that, in the case $\gamma>1$, $v_{\rm a}$ is not the velocity of the car in
  front, but the coarse-grained velocity at a point $x_{\rm a}$ in front of the
  predecessor. A factor $\gamma>1$ means that one anticipates future velocity
  changes of the car in front. Thus, $\gamma$ is a measure of the experience
  and attentiveness of the drivers.
\bibitem{note-bimodal}
Otherwise one would obtain a bimodal distribution where the two peaks would
  correspond to typical desired velocities of cars and trucks.
\bibitem{note-approxtheta} 
The approximation $(\theta+\theta_{\rm a})/2 \approx \theta_{\rm a}$ 
  leads to quantitative
  changes of dynamic properties of the order of 10\%, but simplifies
  the velocity equation considerably. The simpification is justified in the
  dimensionless equations, because they are mainly used for fundamental
  investigations of instability properties and not for a quantitative 
  comparison with empirical data.
\end{references}


\clearpage

\newcommand{\entry}[3]{\parbox{45mm}{#1} &
                      \parbox{12mm}{#2} &
                      \parbox{40mm}{#3} }
\begin{table}

\begin{tabular}{l|l|l}
\entry{Parameter} 
      {Symbol}
      {Typical Value}                  
         \\[0mm] \hline
\entry{Desired Velocity}
      {$V_0$}
      {110 km/h}
         \\[0mm]
\entry{Maximum Density}
      {$\rhomax$}
      {160 vehicles/km}
         \\[0mm]
\entry{Acceleration Relaxation Time}
      {$\tau$}
      {35 s}
         \\[0mm]
\entry{Time Headway}
      {$T$}
      {1.8 s}
         \\[0mm]
\entry{Anticipation Factor}
      {$\gamma$}
      {1.2}
         \\[0mm]
\end{tabular}

\vspace*{10mm}

\caption[]{
\label{tab_par}
Typical parameter values of the GKT model
used for the simulations throughout this paper. The values were obtained by
calibration of the model parameters to Dutch freeway data.}
\end{table}


\unitlength=0.8mm 

\unitlength=0.9mm

\unitlength=0.8mm

\newpage

\unitlength=0.85mm 


\end{document}